\documentclass[]{agujournal2019}
\usepackage{url} 
\usepackage{lineno}
\usepackage[inline]{trackchanges} 
\usepackage{soul}

\usepackage{graphicx,import,dcolumn,float}
\usepackage{threeparttable}
\usepackage{amsmath, amssymb}

\draftfalse

\journalname{Radio Science}

\begin{document}

\title{An Instrument Error Budget for Space-Based Absolute Flux Measurements of the Sky Synchrotron Spectrum Below 20~MHz}


\authors{J. Rolla\affil{1}\thanks{©2023. California Institute of Technology. Government sponsorship acknowledged.}, A. Romero-Wolf\affil{1}, T.J.W. Lazio\affil{1}}

\affiliation{1}{Jet Propulsion Laboratory, California Institute of Technology, 4800 Oak Grove Blvd, Pasadena, CA 91109}

\correspondingauthor{Julie Rolla}{julie.a.rolla@jpl.nasa.gov}




\begin{keypoints}
\item An instrumental error budget was built for space-based measurements of the  absolute flux of the sky synchrotron spectrum below 20~MHz.
\item Contributions from dipole dimensions, plasma noise, stray capacitance, and the amplifier were combined with a Monte Carlo uncertainty model.
\item The impact of individual parameters was explored, highlighting the importance of managing stray capacitance and dipole length uncertainties.
\end{keypoints}

\begin{abstract}
This work describes the instrumental error budget for space-based measurements of the  absolute flux of the sky synchrotron spectrum at frequencies below the ionospheric cutoff ($\lesssim$20~MHz). We focus on an architecture using electrically short dipoles onboard a small satellite. The error budget combines the contributions of the dipole dimensions, plasma noise, stray capacitance, and front-end amplifier input impedance. We treat the errors using both a Monte Carlo error propagation model and an analytical method. This error budget can be applied to a variety of experiments and used to ultimately improve the sensing capabilities of space-based electrically short dipole instruments. The impact of individual uncertainty components, particularly stray capacitance, is explored in more detail.
\end{abstract}


%
%
\section{Introduction}\label{sec:intro}
The rich history of space-based radio astronomical instruments \cite<i.e.>[]{1963SSRv....2..175L,1964AnAp...27..823H,1964Natur.203..173H,1968AJS....73Q..62H,1969JRASC..63R..94H,1970ApJ...160..293H,1970PASP...82R1026A,1971RaSc....6.1085W,1971ITIM...20...86Y,1973ApJ...180..359B,1975RF.....19..250W,1975A&A....40..365A,1977SSRv...21..309W,1978ITGE...16..199K,1978ITGE...16..191S,1992SSRv...60..341G,1995SSRv...71..231B,2001A&A...372..663M,2004SSRv..114..395G,2008SSRv..136..487B,2008SSRv..136..529B,2016SSRv..204...49B,2017SSRv..213..347K,2020A&A...642A..12M}, spans from some of the first missions in the early 1960s to current missions like SunRISE~\cite{Kasper_2022}. The primary focus of many of these instruments was either solar or planetary radio emissions, though some of the instruments were used to characterize the distribution of the Galactic synchrotron emission resulting from relativistic electrons spiraling in the Galaxy's magnetic field.  
Notable among these missions is Radio Astronomy Explorer-2~\cite{1975A&A....40..365A}, as its scientific objectives included studies of Galactic and extragalactic sources.

At frequencies below about 100~MHz, space-based radio astronomy becomes particularly well motivated because of the increasingly strong absorption by the Earth's ionosphere. In the ``high frequency'' (HF) band (3\,MHz -- 30\,MHz), the optical depth of the Earth's ionosphere can exceed unity, rendering it optically thick and precluding observations from the ground. 
The exact frequency depends upon the time of day, the phase of the solar cycle, and the magnetic latitude of the observer. 

However, in this HF band, there are compelling scientific motivations for continued space-based radio astronomical observations -- including the potential of radio emissions from extrasolar planets -- analogous to that observed from Solar System planets~\cite <i.e.>[]{1999JGR...10414025F,2001Ap&SS.277..293Z,2006pre6.conf..603L}, probing the distribution of cosmic ray electrons and cosmic magnetic fields, in concert with X-ray observations~\cite <i.e.>[]{1948PhRv...73..449F,1966ApJ...146..686F,1967MNRAS.137..429R,1970RvMP...42..237B,1996IAUS..175..333F}, and the highly redshifted hyperfine line of neutral hydrogen~\cite{2006PhR...433..181F,2008PhRvD..78j3511P,2010PhRvD..82b3006P,2013PhRvD..87d3002L}.
In contrast to previous radio astronomical observations from space, the expected spectral flux density or brightness for these observations is often well below the synchrotron emission from the Galaxy's disk or extra-galactic synchrotron emission. For simplicity, we refer to this contribution henceforth as the Galactic synchroton emission or Galactic flux.  As such, the Galactic synchrotron emission becomes an irreducible contribution to the sensitivity, or system noise temperature, of a space-based radio astronomical instrument over much of the relevant frequency range (frequencies $\nu \gtrsim 0.5\,\mathrm{MHz}$).
Of particular concern for some proposed measurements of the highly redshifted hyperfine line of neutral hydrogen is that they are \emph{spectral}, so not only is precision flux density or brightness (amplitude) calibration required, but precision spectral calibration as well.

Thus, it is imperative to have a strong understanding of the error budget of the measurement systems and a focus to constrain and characterize that uncertainty. 
While there are several missions that have achieved absolute flux calibrations at higher radio frequencies, such as COBE-FIRAS~\cite{cobe-firas-main, FIRAS_Calibration, FIRAS_Calibration2}, WMAP~\cite{WMAP-2013}, and Planck~\cite{Planck-2020}, these instruments have relied on aperture-filling thermal loads to calibrate the antenna.  Due to the long wavelengths of interest in this study ($>15\,\mathrm{m}$), aperture-filling methods with temperature-controlled loads required is not practical. Instead, we rely on knowledge of the electrically short dipole antenna dimensions and stray capacitance, which determine the conversion from an incident electric field to a voltage at the terminals of the antenna. The internal radiometric calibration using noise diodes used in previous missions can be applied for calibration downstream of the antenna terminals. 

This paper presents an error budget model for the absolute flux measurement uncertainty of a single spacecraft above geosynchronous equatorial orbit (GEO) or above with an electrically short dipole. Section~\ref{sec:sky_below_20MHz} provides a summary of the state of the art of observations and sky models at low frequencies, which are a critical component of estimating the performance of future missions. In Section~\ref{sec:system}, we describe the assumed measurement system featuring a radio antenna and associated electronics for receiving and processing the signals mounted on a spacecraft. The methodology of the error budget, including Monte Carlo uncertainty propagation is described in Section~\ref{sec:methodology}. Sections~\ref{sec:results} and \ref{sec:discussion} present and discuss the initial results. The conclusion provides implications and discusses future improvements to the model.  

\section{The Sky at Frequencies Below 20 MHz} \label{sec:sky_below_20MHz}
\subsection{Observations}\label{sec:observations}

Measurements at frequencies below~20~MHz are challenging due to the opacity of the Earth's ionosphere. Some ground-based measurements exist from times when solar activity was sufficiently low to observe through the ionosphere in this band. Sky spectra were obtained from Hobart down to 1~MHz~\cite{1964Natur.204..171E}. Partial sky images were achieved down to 4.7~MHz from Tasmania~\cite{1966ApJ...143..227E}, including flux measurements for \hbox{Cen~A}, \hbox{For~A}, \hbox{Pic~A}, and \hbox{Hyd~A}. Partial sky images have also been obtained at 10~MHz from the Dominion Radio Astrophysical Observatory (DRAO) in British Columbia, Canada~\cite{1976MNRAS.177..601C}. The Tasmanian low-frequency antenna was able to map the southern sky at 9~MHz with resolution comparable with that of the DRAO results~\cite{1975PASA....2..330C}. The calibrated sky spectral measurements of~\cite{1979MNRAS.189..465C} were down to 5.2~MHz.

Space-based measurements at lower frequencies have been obtained by the Interplanetary Monitoring Platform-6 (IMP-6, 0.130~MHz--2.60~MHz, \cite{Brown_1973}), Radio Astronomy Explorer-1 (RAE-1, 0.4~MHz --6.5~MHz, \cite{Alexander_1969}) and RAE-2 (0.25~MHz--9.18~MHz, \cite{Novaco_1978}). These measurements, along with several ground-based measurements that included absolute flux calibrations are incorporated into the parameterization of \citeA{cane-1979}. While space-based observations have been of significantly lower resolution compared to ground-based observations, these space-based observations are used as a reference for calibration of subsequent space-based radio observatories (e.g., \cite{Dulk_2001, Zarka_2004,Zaslavsky_2011RaSc,Page_2022, Bassett_2023}). The uncertainties on the data below~10~MHz are quoted at $\sim 20\%$ (RAE-2~\cite{Novaco_1978}) but the spread in points can be as high as factors of~2 (see compilation in~\citeA{cane-1979}). A number of experiments for space-based or lunar observations of the low frequency sky are in development~\cite{nature_overview}. For example the The Lunar Surface Electromagnetics Experiment (LuSEE Night) aims to measure radio signals below 50\,MHz from the far side of the moon, using two orthogonal 6\,m full length dipoles~\cite{LuSEE2023}.

\begin{figure}[ht]
    \centering
    \includegraphics[width=1\linewidth]{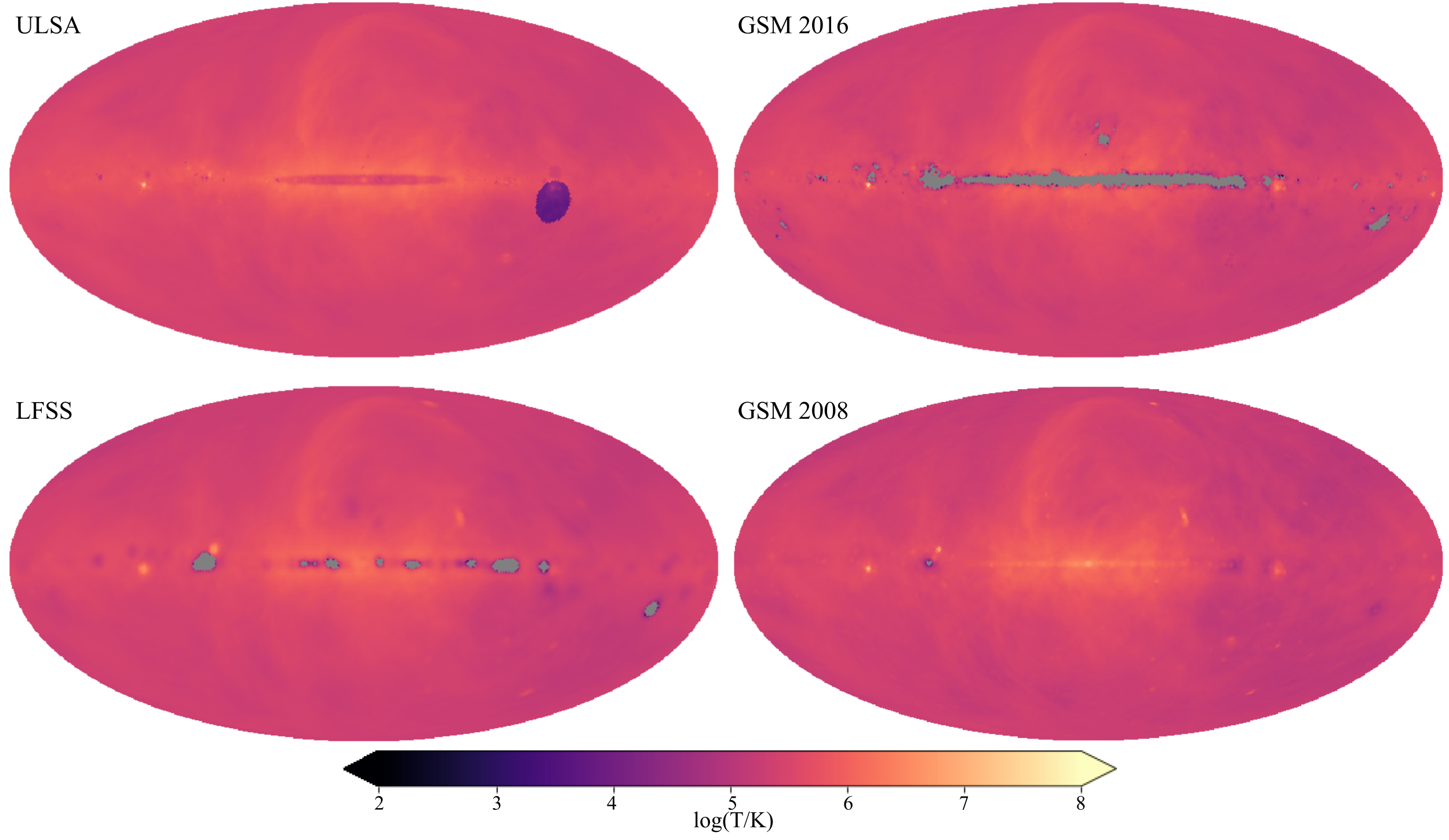}
    \caption{Comparison of 4 different sky maps at 10\,MHz. Top left: USLA~\cite{ULSA} includes the NE2001 sky absorption model~\cite{NE2001}. Top right: GSM 2016~\cite{GSM2016} does not include a low-frequency sky absorption model and a significant portion of the Galactic plane has negative fluxes (shown in gray due to the logarithmic color scale). Similarly, LFSS~\cite{LFSM} (bottom left) and GSM 2008~\cite{GSM2008} (bottom right), lack absorption models and produce negative fluxes near the Galactic plane, although to a lesser extent than GSM 2016.}
    \label{fig:10mhz_maps}
\end{figure}

\subsection{Models}\label{sec:all_sky_models}
There is currently limited data apropos low frequency sky maps, which is critical in understanding the relevant signals of interest. It has previously been common to extrapolate the 408\,MHz Haslam map~\cite{haslam408} to lower frequencies; however, this fails to account for low frequency effects and variations, most notably free-free absorption due to thermal electrons in the Galaxy's interstellar medium \cite{1990LNP...362..121R}.
Table~\ref{tab:skymodels} lists a few efforts to improve upon this method with sky maps created using additional data sources, including the 2008 and 2016 Global Sky Model (GSM)~\cite{GSM2008, GSM2016}, the Ultra Long Wavelength Skymap with Absorption effect (ULSA)~\cite{ULSA}, and the Low Frequency Sky Model (LFSM)~\cite{LFSM}. Of particular note is the limited amount of data below 20\,MHz. Despite the many observations listed in \S\ref{sec:observations}, the only data considered of sufficient quality for GSM and LFSM is the 10~MHz survey of \citeA{1976MNRAS.177..601C} with a $2.9^\circ\times1.9^\circ$ beam resolution (FWHM) and $\sim25\%$ sky coverage. The next dataset considered of sufficient quality for GSM and LFSM is at 22~MHz \cite{Rogers_1999} with a $1.1^\circ\times1.7^\circ$ beam resolution (FWHM) and $\sim35\%$ sky coverage. Another concern is that each of these maps do not incorporate any spatial component to the spectral index~\cite{Spinelli_2021}. This analysis highlights the need for new and expanded sky surveys below 20\,MHz that could significantly improve these models. A comparison of each model at 10\,MHz is shown in Figure~\ref{fig:10mhz_maps}.

\begin{table}
\centering
\begin{threeparttable}[ht]
\caption{Comparison of All-Sky Models}\label{tab:skymodels}
\begin{tabular}{|c|c|c|c|c|c|}
\hline
Model & Year & \begin{tabular}[c]{@{}c@{}}Minimum Frequency\\ (MHz)\end{tabular} & Number of Surveys & \begin{tabular}[c]{@{}c@{}}Angular\\ Resolution\end{tabular} & Method \\ \hline
GSM  & 2008 & 10 & 11       & 5.1$^{\circ}$     & PCA \\ \hline
GSM  & 2016 & 10 & 29       &  5$^{\circ}$      & PCA \\ \hline
LFSM & 2016 & 10 & 20       &  5.1$^{\circ}$    & PCA \\ \hline
ULSA & 2021 & 1  & 10\tnote{*}   &  5$^{\circ}$      & Extrapolation \\ \hline
\end{tabular}
\begin{tablenotes}
\item[*] The 10 surveys used in ULSA are used to produce the spectral index of the power-law and not for an interpolation between the surveys.
\end{tablenotes}
\end{threeparttable} 
\end{table}

The GSM~2008 model was produced using 11 different sky surveys ranging from 10\,MHz to~94\,GHz. The four lowest frequency are 10, 22, 45, and~408\,MHz. The model is built using a Principal Component Analysis (PCA), which determines multiple linear components to explain the trends in the sky maps. The model ultimately determines 3 components are sufficient to explain the majority of the variations in survey data. The authors estimate the error of their model by iteratively predicting one survey using the remaining 10. Predicting the 10\,MHz survey gives an uncertainty of about 9\%, with that number presumably improved if the survey is included.  This map at low frequencies reports negative fluxes (shown in gray) for a region in the Galactic plane (see Figure~\ref{fig:10mhz_maps}), which are unphysical. 

The GSM 2016 model improves upon by the 2008 version by including an additional 18 surveys, bringing the total to 29. This model similarly uses a PCA analysis but uses 5 components instead of 3, corresponding to synchrotron, free-free, cold dust, warm dust, and the CMB anisotropy. Despite these changes, the authors give a similar uncertainty in the map of around 10\% citing that significant extrapolation occurs at the boundaries of the model. This is exemplified by the fact that of the 18 new surveys, the lowest frequencies are 85\,MHz, and 150\,MHz, with the majority of the new maps above 100\,GHz. At low frequencies the extrapolation is dominated by only two components, with the other three virtually negligible. Finally, the extrapolation into lower frequencies produces a significant region of negative brightness, particularly around the Galactic plane and nearby point sources. 

The LFSM is also a PCA analysis based on GSM~2008; however, it incorporates 9 additional maps from the Long Wavelength Array (LWA1) ranging from 35\,MHz, to 80\,MHz. Like GSM 2008, the LFSM finds 3 components in the PCA analysis. While there is no information about the uncertainty at 10\,MHz, at 74\,MHz, the authors give the overall difference of about 10\%, with some regions above 20\%. Note that this map also produces negative brightness temperatures, mostly associated with the Galactic plane that are more significant than the GSM~2008 but less significant than GSM~2016. 

ULSA is the only model discussed here that produces maps below 10\,MHz and takes a different approach. At frequencies below 10\,MHz, the free-free absorption of thermal electrons becomes increasingly significant. ULSA includes free-free absorption by incorporating the NE2001 electron density model~\cite{NE2001}. The ULSA model is generated not by a PCA analysis, but instead by extrapolating from the 408\,MHz Haslam map using a power law, and then using the NE2001 model to add the absorption effect. They also manually add specific small scale dense HII regions. The authors do not investigate the uncertainty or errors of the produced maps. 

\begin{figure}[ht]
    \centering
    \includegraphics[width=1\linewidth]{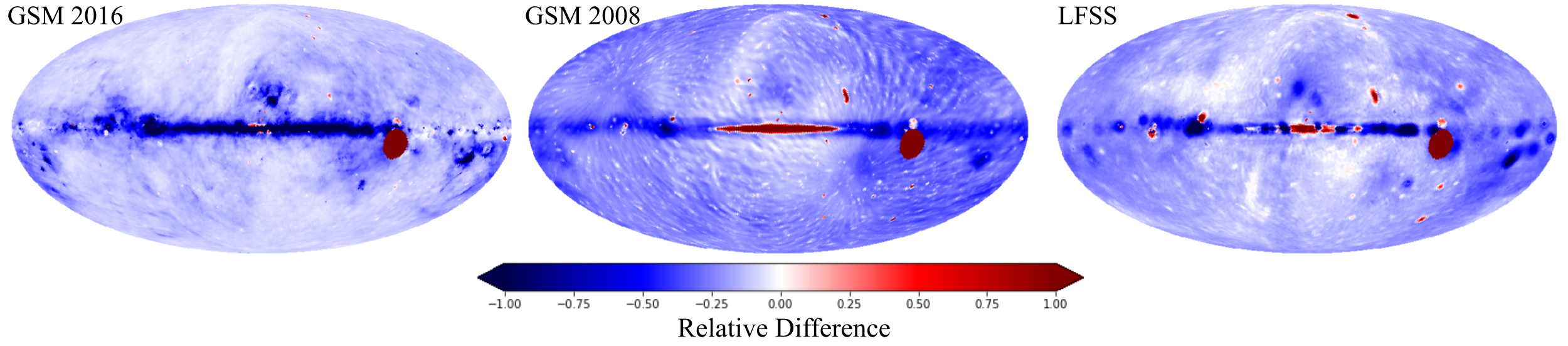}
    \caption{Comparison of the ULSA sky map to the GSM 2016 (left), GSM 2008 (center) and LFSS (right) at 10\,MHz. The gray regions, mostly associated with the Galactic center, are due to negative brightness being reported by GSM 2008, GSM 2016, and LFSS, but not by \hbox{ULSA}. The large red spot exceeding 10\% differences with respect to ULSA are due to the Gum nebula and the fact that ULSA is the only map that includes a free-electron absorption model. Relative difference is given by $(A-B)/B$, calculated at each pixel.}
    \label{fig:rel_diff_maps}
\end{figure}

Figure~\ref{fig:rel_diff_maps} shows relative difference maps to compare ULSA at 10\,MHz to each of the other models. ULSA tends to have a higher temperature throughout the map, except for specific point sources and near the Galactic center. 

\begin{figure*}[t]
    \centering
    \includegraphics[width=1\linewidth]{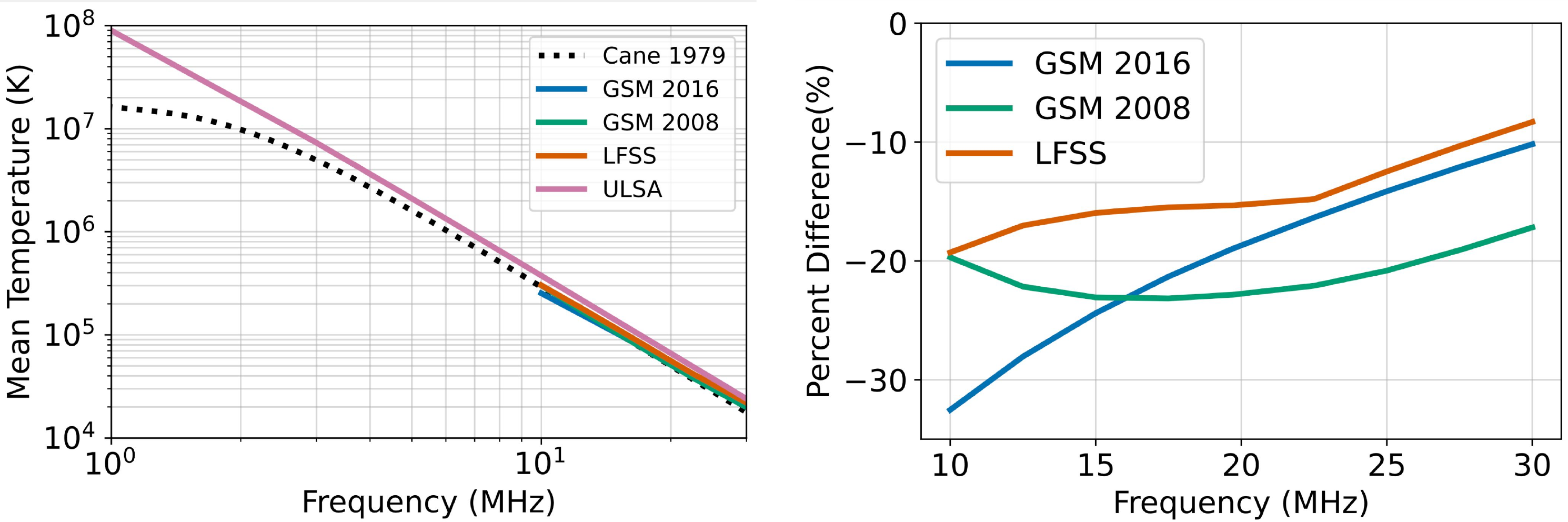}
\caption{Left: Sky-averaged temperature of each model at frequencies below 30\,MHz. Note that while ULSA goes below 1\,MHz, the other models minimum frequency is 10\,MHz. Included for reference is the~\cite{cane-1979} parameterization of the sky brightness, converted to temperature. While the ULSA model appears linear, there is slight kink at 1\,MHz, which is more apparent when converted to brightness (as in the ULSA publication). Right: Percent differences between the ULSA sky-averaged temperature and the other models.}
    \label{fig:meantemps}
\end{figure*}

The left panel of Figure~\ref{fig:meantemps} shows the sky-averaged temperature of each model at frequencies up to 30\,MHz. The trends at 10\,MHz described above are continued at higher frequencies with discrepancies between these models ranging between 10\% and 30\%, depending on the frequency (see right panel of Figure~\ref{fig:meantemps}). Note that GSM 2008, GSM 2016, and LFSS do not match the observed turnover in the spectrum due to Synchrotron Self Absorption in the Cane parametrization but ULSA does. 

Below 10~MHz we can compare ULSA to the Cane parameterization of the sky brightness (see Eq.~\ref{eq:cane}) using a conversion to temperature
\begin{equation} 
T_f = \frac{I_f c^2}{2k_Bf^2}
\end{equation} 
where $T_f$ and $I_f$ are the temperature and brightness respectively at frequency, $f$. The left panel of Figure~\ref{fig:meantemps} shows that ULSA does not reproduce the spectral break at $\sim2$~MHz of Cane 1979, as recognized by the authors of the model~\cite{ULSA}.

In addition to these discrepancies, none of these models provide uncertainties. The large differences of these models necessitates a robust understanding of an experiment's error budget in order to utilize the sky models in the measurement of potential science signals of interest.


\section{Measurement System Model}\label{sec:system}
\subsection{System Overview}

\begin{figure}[ht]
    \centering
    \includegraphics[width=1\linewidth]{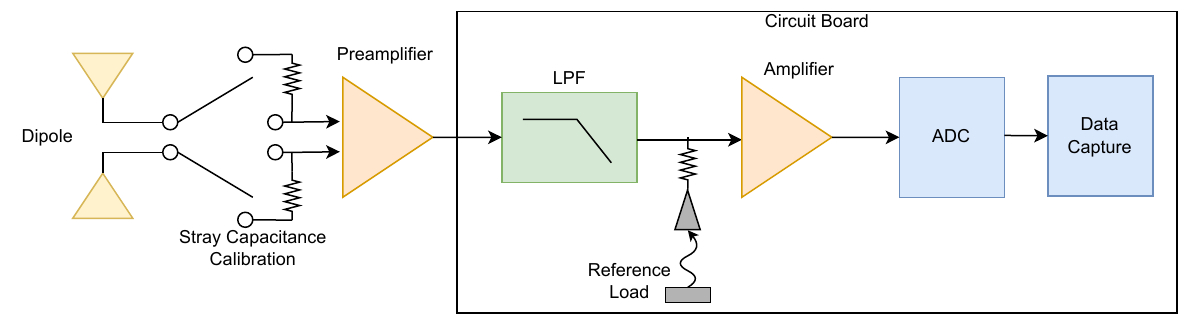}
    \caption{Measurement system, consisting of an antenna and associated electronics for processing the received signals. This includes a system to monitor and calibrate stray capacitance and a reference thermal load.}
    \label{fig:system}
\end{figure}

Figure~\ref{fig:system} presents an instrument diagram of the measurement system. This includes a dipole antenna, a stray capacitance calibration system, low-pass filters (LPF), a reference load, amplifiers, an analog-to-digital conversion (ADC), and a data capture system. This system is designed to measure the absolute flux across a frequency band below~20\,MHz. The approach of this paper is to consider an electric field incident on the electrically short dipole antenna and concluding with data recorded by the spacecraft.

The absolute-flux calibration for this system cannot be achieved with traditional aperture-filling thermal radiators given the wavelengths and large field of view of the antenna. To achieve this, we instead include a monitor the stray capacitance of the system (\S\ref{sec:stray_capacitance}) and include a reference thermal load. The signal then passes through a first stage amplifier and analog-to-digital converter (ADC) before being saved and prepared for data transfer. An intermediate stage of signal conditioning between the first-stage amplifier and the ADC can also be included (e.g., second-stage amplifiers and filters) but its design typically depends on the choice of parts used in implementation and does not drive absolute flux calibration requirements.

\begin{figure}[ht]
    \centering
    \includegraphics[width=.9\linewidth]{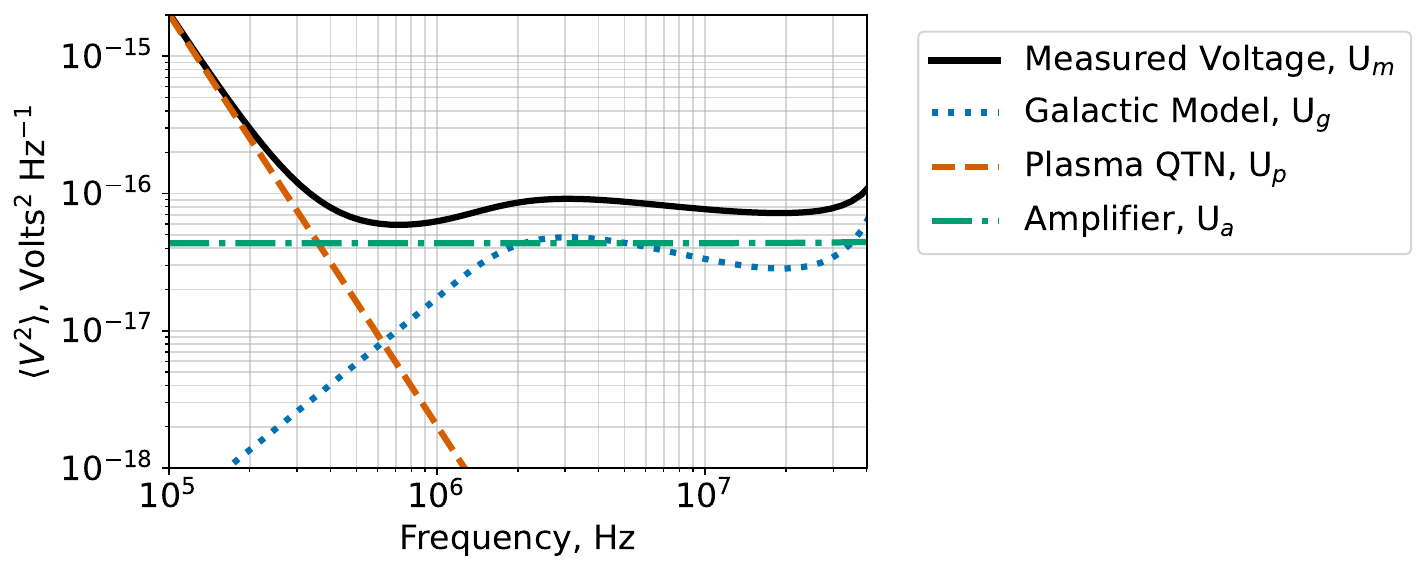}
    \caption{The assumed voltage measured by the detector with a 3\,m dipole is shown in black, given by the sum of the three components: Galactic Model ($U_g$ - blue dotted), Plasma Quasi-Thermal Noise ($U_P$ - orange dashed), and Amplifier Noise ($U_a$ - green dash-dotted). The parameters used can be found in Figure~\ref{fig:errorbudget}.}
    \label{fig:measuredvoltage}
\end{figure}

Given that there may be diverse science motivations, we illustrate the antenna impedance as a function of length at frequencies chosen to span a reasonable set of science motivations from 0.1\,MHz to 25\,MHz. 
In this analysis, we model the voltage power spectral density (PSD) measurement of the receiver, in units of Volts$^2$/Hz, across this frequency range by combining three signal or noise sources:
\begin{equation} \label{eq:measuredvoltage}
U_m = U_g + U_p + U_a
\end{equation}
where $U_g$ is the voltage PSD resulting from the absolute Galactic flux, $U_p$ is the voltage PSD from the Plasma Quasi-Thermal Noise (QTN), $U_a$ is the voltage PSD introduced by the amplifier. The resulting signal and its three components are shown in Figure~\ref{fig:measuredvoltage}. The parameters used to derived the figure are shown in Figure~\ref{fig:errorbudget}. The goal of this analysis is to isolate the Galactic flux, $S_g$, through the equation

\begin{equation} \label{eq:galactic_flux_conversion}
U_g = \chi \times S_g
\end{equation}
where ${\chi}$ is the flux to voltage PSD conversion term described in \S\ref{sec:sky_flux}. 

The Plasma QTN noise, described in \S\ref{sec:PQTN}, is a voltage PSD that occurs at the terminals of antenna and is a function of the electron density, electron temperature and dipole length. The last component, the amplifier noise $U_p$, assumes a operational amplifier which has noise contributions from internal voltage noise, internal current noise and Johnson-Nyquist thermal noise. The amplifier noise will be covered in \S\ref{sec:amp_noise}.

\subsection{Sky Flux}\label{sec:sky_flux}
\begin{figure}[ht]
    \centering
    \includegraphics[width=.6\linewidth]{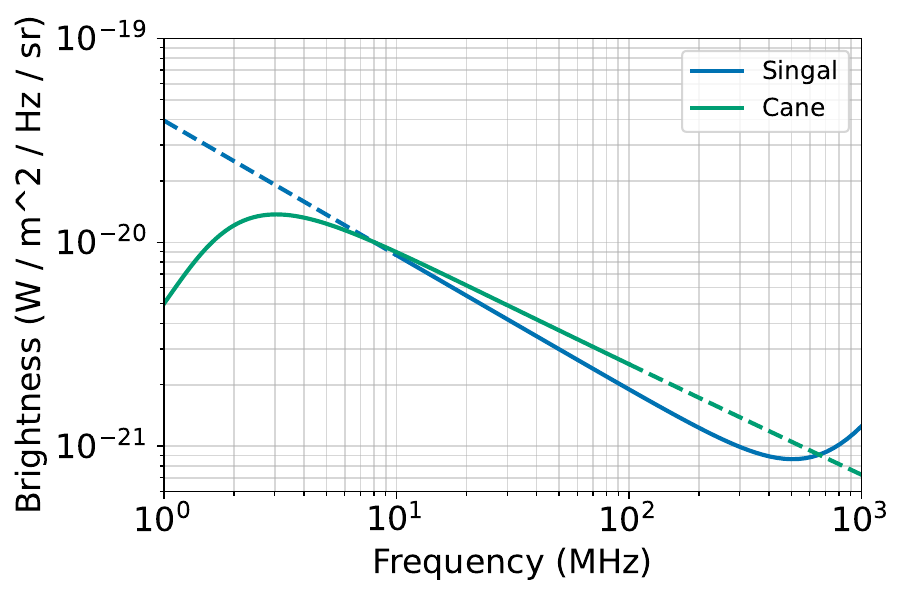}
    \caption{Comparison of sky brightness paramertizations from~\citeA{cane-1979} and~\citeA{Dowell_2018, Singal_2023}. Required extrapolation is shown by the dotted lines.}
    \label{fig:cane_comparison}
\end{figure}

The sky flux (Galactic and extragalactic) is what we aim to measure. For the purposes of this study, we use the parametrization of Cane~(1979) to represent the sky flux.
This parameterization was chosen because it incorporates measurements below 10\,MHz, whereas other more recent parameterizations would require extrapolation into the region of interest~\cite{Dowell_2018, Singal_2023}. Dowell and Taylor~(2018) sky brightness values are within $\sim10\%$ of Cane~(1979) between 5\,MHz and 25\,MHz, as illustrated in Figure~\ref{fig:cane_comparison}.

The parameterization of Cane~(1979) is given by
\begin{equation}\label{eq:cane}
S_g = 4\pi \left[I_g\, f^{-0.52}\,\frac{1-e^{-\tau}}{\tau} + I_{eg}\, f^{-0.8} \,e^{-\tau}\right] \frac{\mathrm{W}}{\mathrm{m}^2\,\mathrm{Hz}\,\mathrm{sr}},
\end{equation}
where the brightness from the Galaxy is given by $I_g = 2.48\times 10^{-20}\,\mathrm{W}\,\mathrm{m}^{-2}\,\mathrm{Hz}^{-1}\,\mathrm{sr}^{-1}$, the brightness from extragalctic sources is $I_{eg} = 1.06\times 10^{-20}\,\mathrm{W}\,\mathrm{m}^{-2}\,\mathrm{Hz}^{-1}\,\mathrm{sr}^{-1}$, and the optical depth is $\tau = 5.0 f^{-2.1}$.
The $4\pi$ term is added to convert the brightness in Cane, to the flux using $ S = B \int d\Omega = B 4\pi$, where $\Omega$ is the solid angle. The result, as a voltage \hbox{PSD}, is shown in Figure~\ref{fig:measuredvoltage} as the blue dotted line.

The voltage PSD from the Galactic sky brightness at the input to the first stage amplifier can be obtained through the flux to voltage conversion $\chi$ given by 
\begin{equation}
{\chi} =   \frac{R_{rad}\,\lambda^2}{\pi} \Gamma^2
\label{eqn:voltage_to_flux},
\end{equation}
where $\lambda$ is the wavelength, $R_{rad}$ is the radiative resistance (\S\ref{sec:antenna}) and $\Gamma$ is the voltage divider, which is dominated by stray capacitance, is discussed in \S\ref{sec:vd}.
\ref{A:flux to voltage} provides a full derivation of the flux-to-voltage PSD conversion.

\subsection{Plasma Quasi-Thermal Noise}\label{sec:PQTN}

Plasma Quasi-Thermal noise is produced by the thermal motion of ambient electrons. We use the parametric approach developed by~\citeA{Meyer-Vernet_1989}. The voltage induced on the antenna is dependent on the temperature of the plasma ($T_e$), the density of electrons ($n_e$), the signal frequency ($f$), and the dipole half length ($L_m$). 

\begin{equation} 
U_p = 5\times10^{-5} \ \mathrm{\frac{V^2}{Hz}} \ 
\left(\frac{n_e}{\mathrm{ cm^{-3}}}\right) 
\left(\frac{T_e}{\mathrm{K}}\right) 
\left(\frac{f}{\mathrm{Hz}}\right)^{-3} 
\left(\frac{L_m}{\mathrm m }\right)^{-1} 
\end{equation} 

As this is an instrument error budget, only the length has an uncertainty. For the assumed orbit above GEO, we assume a constant value of 5\,cm$^{-3}$ for the electron density, and 10,000\,K for the temperature~\cite{plasma_properties}. Lower orbits would have larger electron densities and temperatures that would increase contribution of the plasma quasi-thermal noise, especially at frequencies below 1\,MHz. The electron temperature and density are not well-defined and further investigation into the effects on of different plasma models on this analysis could be investigated in the future. The result is shown in Figure~\ref{fig:measuredvoltage} as the orange dashed line. 

\subsection{Amplifier Noise}\label{sec:amp_noise}

The noise due to the amplifier, shown in Figure~\ref{fig:measuredvoltage}, is the combination of operational amplifier current and voltage noise, as well as the Johnson-Nyquist thermal noise~\cite{nyquist1928thermal}: 
\begin{equation}
U_a = U_{V} + U_{I} + U_{J}
\end{equation} 
The operational amplifier voltage noise \hbox{PSD}, $U_V$, is simply $V_n^2$, where $V_n$ is the voltage noise density specification of the component used. The operational amplifier current noise \hbox{PSD}, $U_{I}$, is $(I_n\times |Z_{eq}|)^2$, where $I_n$ is the current noise density specification of the component used, and $Z_{eq}$ is the impedance of the circuit looking out from the input terminals of the amplifier. Note that the current noise is dependent on the antenna geometry because $Z_{eq}$ is related to the antenna impedance. Finally, the Johnson–Nyquist thermal noise is given by $U_{J} = 4k_B T_0 R_T$, where $k_B$ is the Boltzmann constant, $T_0$ is the temperature of the system, and $R_T$ is the total resistance. As with the current noise, the Johnson-Nyquist noise is dependent on the antenna geometry because the $R_T$ include the radiative resistance. For a more detailed treatment of noise modeling with operational amplifier devices, see~\cite{AD_app_note}. Operation amplifiers have impedance that scales with $f^{-2}$ that can become a significant contributor to noise at low frequencies, but this is typically below our band of interest is therefore not included in this analysis.

The internal noise of the amplifier can vary due to ambient temperature effects. This noise contribution and its variations during flight can be monitored using a reference load for calibration. The Radio Astronomy Explorer employed such a system and achieved an uncertainty of 0.2\,dB ($\sim$0.5\% )~\cite{Weber_1971} on the internal instrument noise contribution. It is expected that a modern system implemented at these frequencies can achieve comparable uncertainties on the internal noise calibration. 

\subsection{Antenna}\label{sec:antenna}

This analysis focuses on a system that uses an electrically short dipole. Dipoles have been proven in various space missions and are shown to be reliable, with STACER deployments being most common. Examples of missions utilizing dipole antennas include STEREO/WAVES~\cite{StereoWaves-antennas}, JUNO~\cite{juno-overview}, JUICE~\cite{JUICE-overview}, and SunRISE~\cite{sunrise-overview-2022}. Operating electrically short dipoles far from resonance enables reception of a larger bandwidth, which may be required by the desired science measurements, and prevents potential stability issues near resonance. Also, short antennas are less susceptible to being damaged or cut by micrometeroids and dust, which occurred to one of the \textit{Wind}/WAVES antennas that was cut twice by dust impacts~\cite{dust_impacts_winds, dust_winds_2}.

Electrically short dipole antennas are used because they are relatively simple systems that can be modeled with few parameters thereby improving the handling of systematic uncertainties. As a dipole approaches resonance, its behavior becomes more complex and can be more susceptible to systematic uncertainties. An important question to address is to what extent is the electrically short dipole approximation valid and what level of uncertainties could one expected from limited knowledge of the antenna parameters.

The antenna impedance is used in a number of calculations required to determine the total uncertainty of the system. The impedance modeled for this analysis uses antennas simulated with Numerical Electromagnetics Code 2 (NEC2)~\cite{NEC2}. Impedance is a complex value that describes the general term for the opposition of a circuit to alternating current given by $Z = R + jX$, where $R$ is the resistance and $X$ is the reactance. In an antenna system, the resistive component can be broken down into a radiative resistance term $R_{rad}$ and an ohmic resistance that results in signal losses $R_{loss}$ such that $R=R_{rad}+R_{loss}$. While the loss resistance can be large, the radiative resistance accounts for the coupling between electric fields incident on the antenna and the voltage produced at the terminals of the antenna via the effective length $h_{eff}\propto R_{rad}$. $R_{loss}$ can be significant for transmitting antennas, for the case of a receiving antenna, the effect is negligible. The reactance for an antenna system does not mirror the resistance terms; instead, it can be described by the capacitive and inductive opposition to a current and is primarily related to the frequency or length. At resonance, the capacitive and inductive components are equal and cancel each other out.

\begin{figure*}[t!]
    \centering
    \includegraphics[width=1\linewidth]{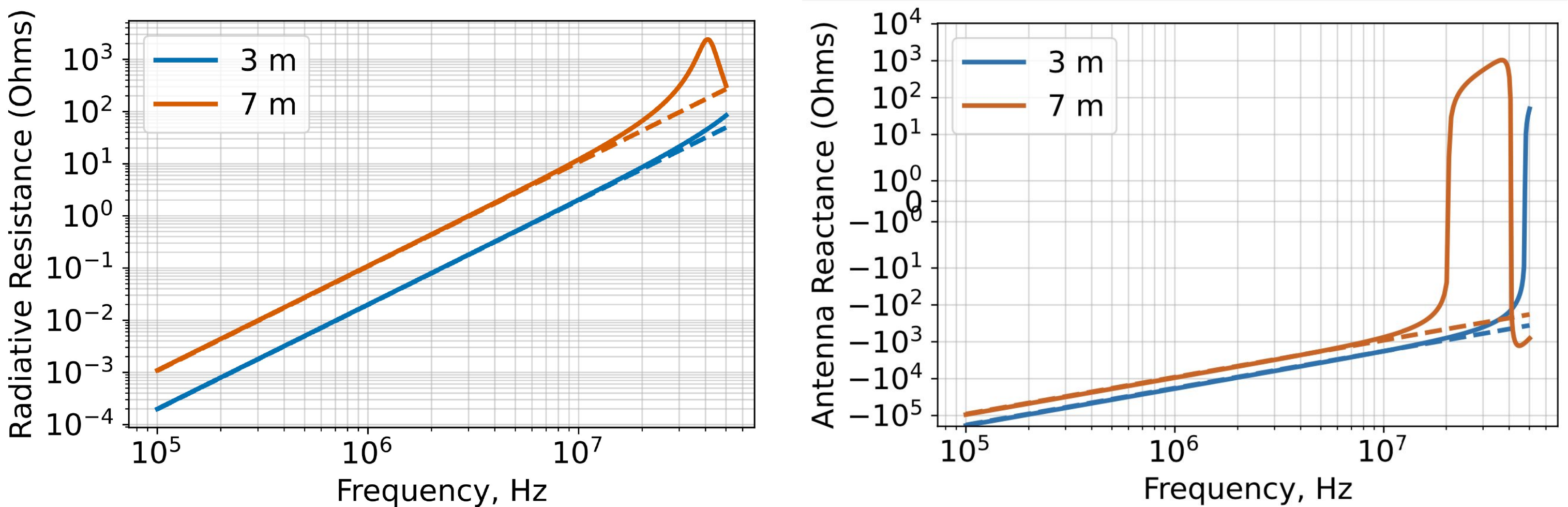}
        \caption{The radiative resistance (left) and antenna reactance (right) of a 3\,m (blue) and 7\,m (orange) dipole. The results of NEC2 simulation is given as a solid line, and the electrically short dipole approximation is shown as a dashed line.}
    \label{fig:antenna_impedance}
\end{figure*}

Figure~\ref{fig:antenna_impedance} (left panel) shows the radiative resistance ($R_{rad}$) of a 3\,m and 7\,m dipole obtained using a NEC2 simulation over the 1-25\,MHz band compared to the electrically short antenna approximation:
\begin{equation} 
R_{rad} \simeq 20\pi^2 \left(\frac{L}{\lambda}\right)^2
\end{equation} 
where $L$ is the full dipole length (tip-to-tip), and $\lambda$ is the wavelength. For the case of a 7\,m antenna, the behavior deviates from the electrically short approximation well before its resonance peak at 20.38\,MHz. For the 3\,m antenna, on the other hand, $R_{rad}$ is well described by the electrically short antenna approximation over the HF radio band.

The cost of using an electrically short dipole, however, is that the antenna reactance takes on large magnitudes, which results in a lower tolerance to stray capacitance (treated in \S\ref{sec:stray_capacitance}).
At frequencies below the first dipole resonance, the antenna reactance can be approximated as a capacitance via 
\begin{equation} 
X_{rad} = \frac{-j}{\omega C_a}
\end{equation} 
where $\omega = 2\pi f$ is the angular frequency, and $C_a$ is the antenna capacitance. For an electrically short dipole, the antenna capacitance can be approximated by
\begin{equation}
C_a \simeq \frac{2\pi \epsilon_0 L_m }{\ln(L_m/a)-1},
\label{eq:antenna_capacitance}
\end{equation}
where $L_m$ is the half length of a dipole antenna, $a$ is the diameter, and $\epsilon_0$ is the permittivity of free space. 
This method is equivalent to using the short dipole approximation of antenna reactance:
\begin{equation} 
X_{rad} = \frac{-120\, \Omega}{\pi L_m/\lambda}\left[\ln{\left(\frac{L_m}{a}\right)} -1 \right]
\end{equation} 

Figure~\ref{fig:antenna_impedance} (right panel) shows the dipole antenna reactance for a 3\,m and 7\,m (tip-to-tip length) dipole, each with a diameter of $a=1$\,cm. As with the radiative resistance, the antenna reactance of a 3\,m dipole is well approximated by a closed-form approximation using Equation~\ref{eq:antenna_capacitance}. For frequencies near resonance, small errors in the length or diameter have bigger impacts on the modeling compared to low frequencies where the electrically short approximations are valid.

\begin{figure*}[t!]
    \centering
    \includegraphics[width=1\linewidth]{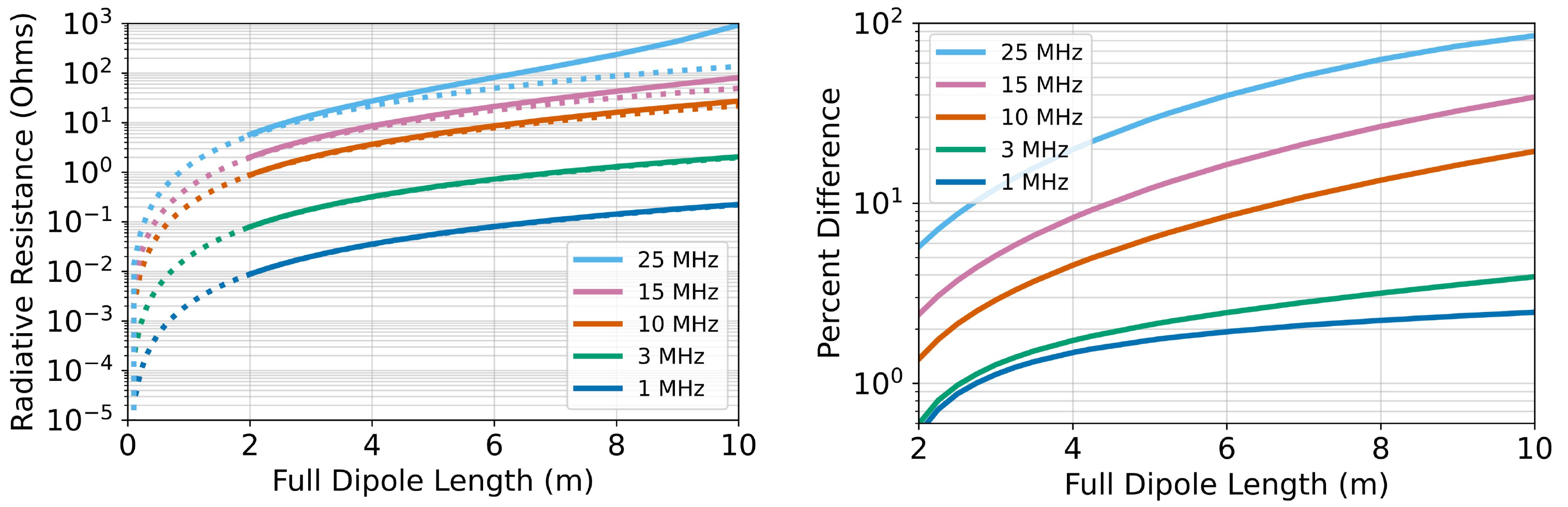}
        \caption{Left: The antenna radiative resistance versus full dipole length at five different frequencies. The NEC2 simulation shown by the solid line and the dotted line shows the short dipole approximation. Right: The percent relative difference between the NEC2 simulation results and the short approximation.}
    \label{fig:resistance_vs_length}
\end{figure*}

\begin{figure*}[t!]
    \centering
    \includegraphics[width=1\linewidth]{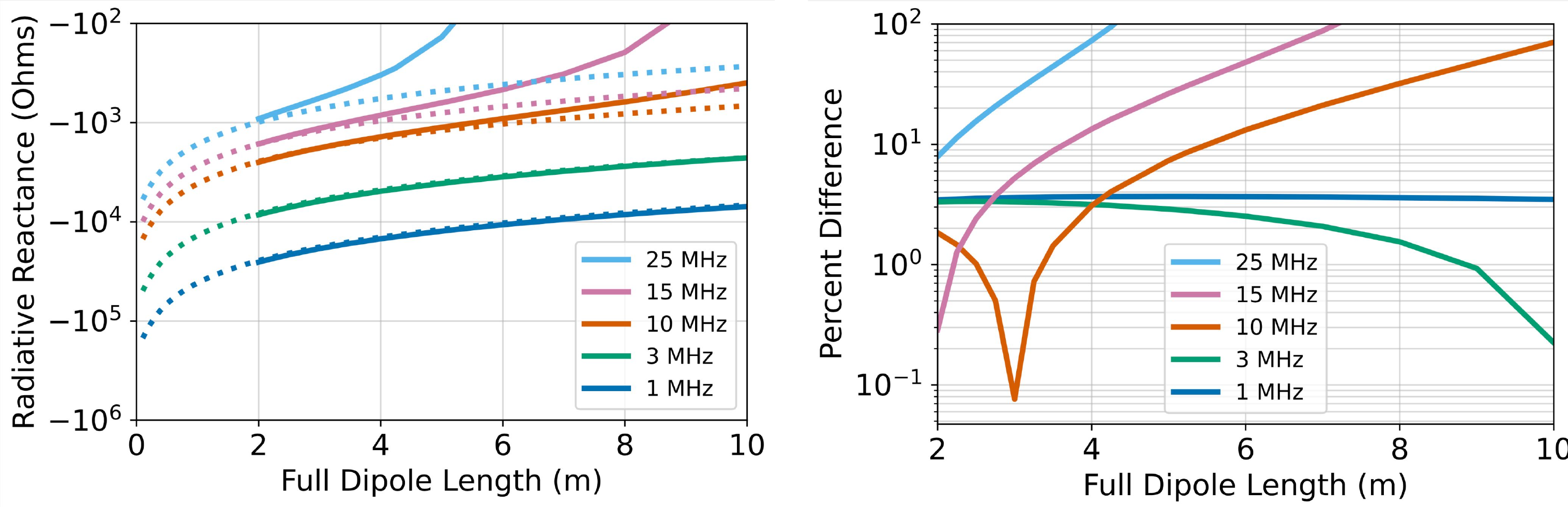}
        \caption{Left: The antenna reactance versus full dipole length at five different frequencies. The NEC2 simulation shown by the solid line and the dotted line shows the short dipole approximation. Right: The percent absolute relative difference between the NEC2 simulation results and the short approximation.}
    \label{fig:reactance_vs_length}
\end{figure*}

We estimate the expected level of systematic uncertainty due to the electrically short approximation by comparing its behavior to NEC2 simulations in Figures~\ref{fig:resistance_vs_length} and~\ref{fig:reactance_vs_length}. The left panel of Figure~\ref{fig:resistance_vs_length} shows the radiative resistance as a function of length for fixed frequencies of 1, 3, 10, 15, and 25\,MHz using NEC2 simulations (solid lines) compared to the electrically short antenna approximation (dotted lines). The right panel of that same figure shows the percent difference between NEC2 and the electrically short approximation. We find that even in the regime where the electrically short antenna approximation is valid, we have a few~\%-level deviations with significant deviations as the electrically short antenna approximation breaks down with increasing length. Similarly, Figure~\ref{fig:reactance_vs_length} shows the reactance as a function of dipole length on the left panel and the deviation from the electrically short dipole approximation on the right panel. The scale of deviation is similar to the comparison of radiative resistance, although the behavior of the curves can vary. 


In \ref{A:finite dipole} we compare the finite dipole analytical solution to the electrically short approximation and NEC2 model. Although the agreement between NEC2 and the finite analytical solution is improved, it does not fully account for the discrepancies.


\subsection{Stray Capacitance Calibration}
\label{sec:stray_capacitance}
Stray capacitance results from the sum total of capacitive couplings between the antennas and the surrounding conductive elements such as the spacecraft chassis, antenna mounts, and any other nearby hardware. The combined effects act as a shunt capacitance in the circuit-equivalent model of the front-end portion of the receiver (i.e., all components between the antenna and first-stage-amplifier). The shunt effect means that stray capacitance ($C_s$) will result in signal losses proportional to $(1+C_s/C_a)^{-2}$ where $C_a$ is the equivalent capacitance of the reactive component of the electrically short dipole antenna impedance.  

The magnitude of the impact of stray capacitance can be minimized by increasing the antenna capacitance $C_a$. The dominant contributor to the antenna capacitance is the antenna length $C_a\propto L$, which is itself limited by the need to keep the length electrically short. While stray capacitance can be measured prior to launch, it is possible that the values shift after launch and antenna deployment and they could vary due to thermal fluctuations. 
Given these considerations, we therefore include a system that can monitor changes in stray capacitance during flight. This could be achieved by switching a known resistive load between the path of the antenna and the first-stage amplifier. The voltage ratios between the measurements with and without the resistive load allow us to measure the stray capacitance in the system. A different but related approach, using an array of inductors, was used for on-board capacitance measurements applied to Radio Astronomy Explorer~\cite{Weber_1971}.


\subsection{Voltage divider}\label{sec:vd}
The process of converting between flux and voltage is an additional main source of uncertainty. The conversion term is dependent on the antenna impedance ($Z_a$), which is dependent on length and diameter, the stray capacitance ($Z_s$), and the load impedance ($Z_L$). These components act as a voltage divider, a dimensionless ratio of impedances that results in a decrease in voltage.
The voltage divider is given by:
\begin{equation} 
\Gamma= \frac{Z_{SL}}{Z_{a} + Z_{SL}}
\end{equation} 
where $Z_{SL}$ is the parallel impedance of the stray and load impedance given by 
\begin{equation} 
Z_{SL} = \frac{1}{\frac{1}{Z_s}+\frac{1}{Z_L}}
\end{equation}

\subsection{Calibration}
Additional calibration terms are needed to interpret the measured voltages at the amplifier, 
including signal chain leakage, amplifier gain, and bandpass response. The amplifier gain and bandpass response are monitored with a reference load (see \S\ref{sec:amp_noise}) where the system is switching back and forth between measuring the load and the antenna receiver signal path. Given we expect $\sim$0.5\% noise calibration errors, we allocate 1\% calibration uncertainties to the gain and bandpass response on-board calibrations. Signal chain leakage arises in systems where more than one polarization is fed into the same receiver, which arise from first, the antenna cross-polarization response and second, internal couplings in the signal chains downstream of the antenna terminals. The cross-polarization response arises from the coupling of signals in the polarization opposite to the antenna. For example, for a dipole oriented along one axis, electric fields incident from different axes can induce a voltage at the terminals. In the case of the signal chain, a dual-polarized antenna will have signal chains for each polarization that are close to each other and there can be reactive couplings between them. The leakage from internal couplings downstream of the antenna is typically much smaller than 1\% so we do not include it here. The antenna cross-polarization is constrained via alignment tolerances and detailed antenna simulations. We allocate 1\% uncertainties for this calibration contribution.


\section{Methodology}\label{sec:methodology}



\begin{figure}
    \centering
    \includegraphics[width=.9\linewidth]{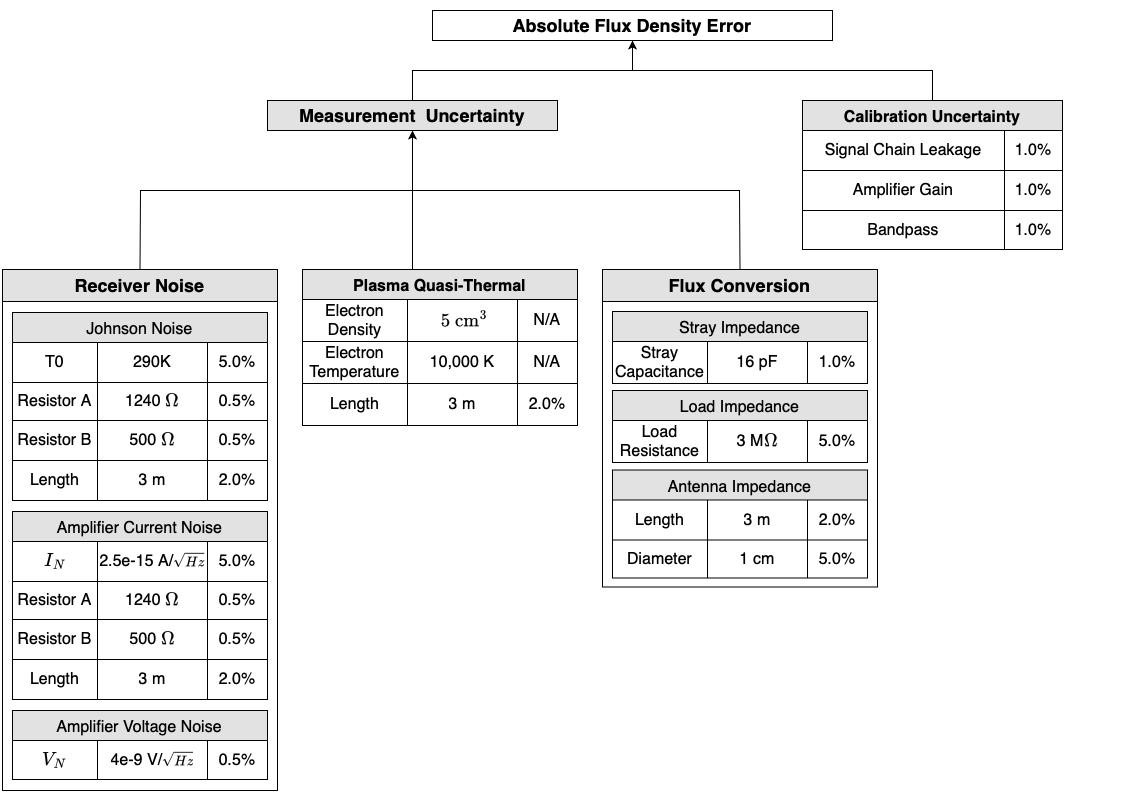}
    \caption{Diagram illustrating sources of uncertainty in determining the background diffuse flux. The three noise signals (amplifier, plasma QTN, and other system noise) and their components are shown. The voltage divider, due to the stray capacitance, introduces significant uncertainties in the conversion from voltage to flux.}
    \label{fig:errorbudget}
\end{figure}

A global error budget for absolute sky flux measurements includes contributions from the instrument, astrophysical sources, and environmental parameters such as variations in the local plasma environment (e.g. quasi-thermal noise) and radio frequency interference. In this paper we focus on the instrumental error budget.
The basic structure of the instrumental error budget, with assumed parameter values and associated percent uncertainties, is presented in Figure~\ref{fig:errorbudget}. The sources of noise can be divided into two categories, Measurement and Calibration Uncertainty. The Measurement Uncertainty is associated with signals that are being detected, while the Calibration Uncertainty is a result of secondary effects of the spacecraft system. 

There are three main sources of measurement uncertainty, each with multiple parameters that feed into them: Plasma Quasi-Thermal Noise, amplifier noise and the flux density to voltage PSD conversion. The flux to voltage PSD conversion uncertainty is dominated by the effect of shunt capacitance which results from the combination of the antenna impedance and stray capacitance. 
The Plasma Quasi-Thermal Noise is a result of the thermal activity of ambient plasma electrons. Note that because the antenna geometry is a component in many of the measurement uncertainty  calculations, the antenna length is present multiple times in the table. The uncertainty on the antenna geometry terms is due to the imperfect deployment of dipole antenna boom elements, not manufacturing tolerances.

The calibration uncertainty is modeled as an additive white Gaussian noise component, with three individual components, each with an assumed uncertainty of 1\%: Signal Chain Leakage ($N_{SC}$), Amplifier Gain($N_{AG}$), and Bandpass response($N_{BR}$). These factors are added in quadrature to get the total calibration uncertainty, $N$:
\begin{equation}
    N = \sqrt{N_{SC}^2 + N_{AG}^2 + N_{BR}^2}
\end{equation} 
The calibration uncertainties propagate to the measured voltage PSD, which subsequently propagates to the estimated Galactic noise measurement. Given the assumed uncertainty of 1\% for each component, $N = 1.7\%$. This produces a variance equal to the measured voltage multiplied by $N$. Finally, the voltage is adjusted through a normal distribution centered on the measured voltage, with the variance described above: 
\begin{equation} 
U_m' = \mathcal{N}(U_m, N U_m)
\end{equation} 
where the prime indicates calibration corrections have been applied to the total measured voltage PSD.

The calculation of the voltage PSD contribution due to that Galactic flux is thus given by
\begin{equation} 
U_g' = U_m' - U_p' - U_a'
\end{equation} 
%
%
The uncertainty on the estimated Galactic flux results from the propagation of $S_g = U_g'/\chi.$


For the uncertainty propagation of the errors in this analysis, both analytical and Monte Carlo approaches are used. By using both methods, we ensure that the calculations are valid and provide understanding of the magnitude of the simplifying assumptions required to perform the analytical calculations. A full derivation of the analytical error propagation of the models presented in Section~\ref{sec:system} 
is provided in \ref{A:analytical}. Monte Carlo uncertainty propagation works by sampling parameter values from a distribution and performing each calculation numerous times. The result of these calculations gives a posterior distribution, from which the uncertainty can be determined. In this analysis, the uncertainty is reported as the percentage of the mean given by one standard deviation of the resulting distribution. Statistical error is assumed to be negligible compared to systematic uncertainties in this analysis but will be the subject of future work. At frequencies below 1\,MHz (outside the scope of this analysis) it would be important to evaluate the required integration times for this assumption as variability in parameters such as the plasma electron density and temperature could make long time scale measurements difficult.

 \begin{figure}[ht]
    \centering
    \includegraphics[width=.9\linewidth]{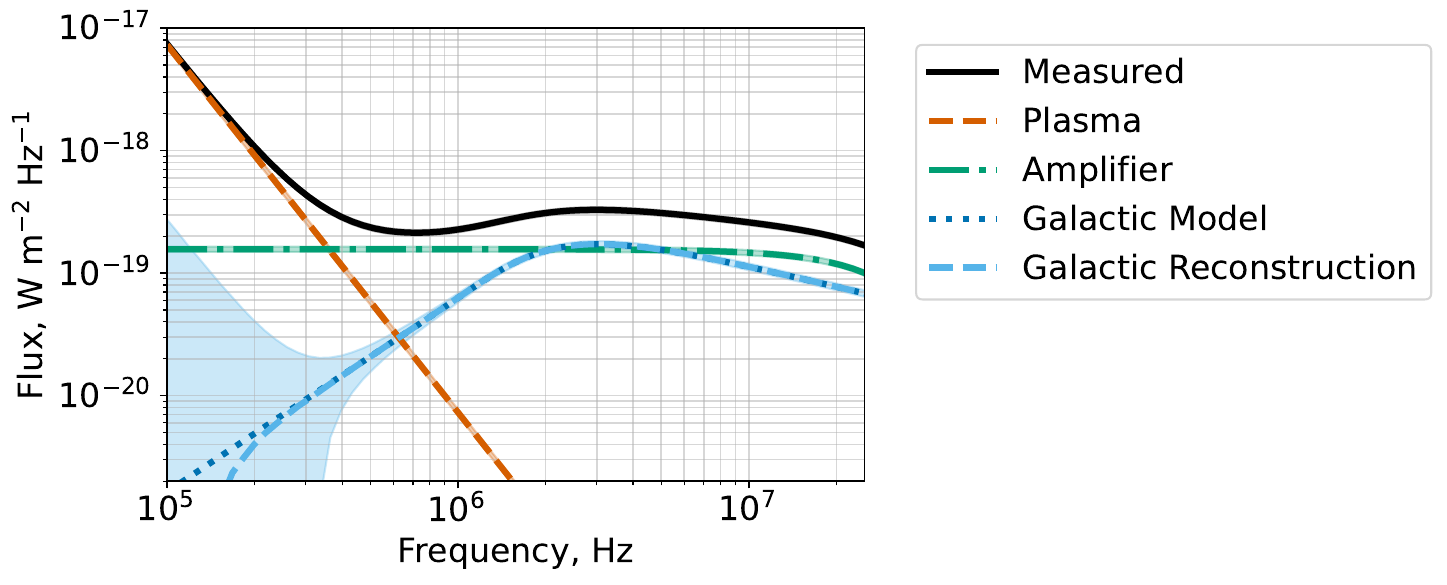}
    \caption{Flux results of the Monte Carlo Uncertainty propagation with NEC2 antenna simulations. The Galactic flux prediction is shown in light blue, with the colored band representing the final uncertainties. Each of the components also has a band showing the uncertainty, but they are not clearly visible. The predicted flux (light blue) is effectively the measured flux (black) minus the plasma flux (orange) and amplifier flux (green).} 
\label{fig:final_flux_uncertainties}
\end{figure}

 \begin{figure}[ht]
    \centering
    \includegraphics[width=.6\linewidth]{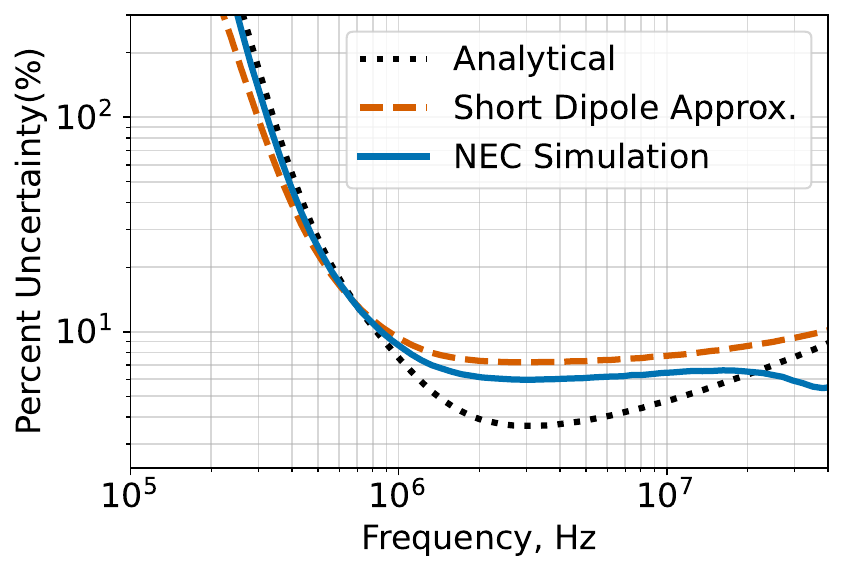}
    \caption{The predicted flux uncertainty, calculated using the analytical equations (black dotted line) from \ref{A:analytical}, Monte Carlo Uncertainty propagation using the short dipole approximation (orange dashed line), and the Monte Carlo Uncertainty propagation using NEC2 simulations (blue solid line).}
    \label{fig:flux_uncertainty}
\end{figure}

\begin{figure*}[t!]
    \centering
    \includegraphics[width=1\linewidth]{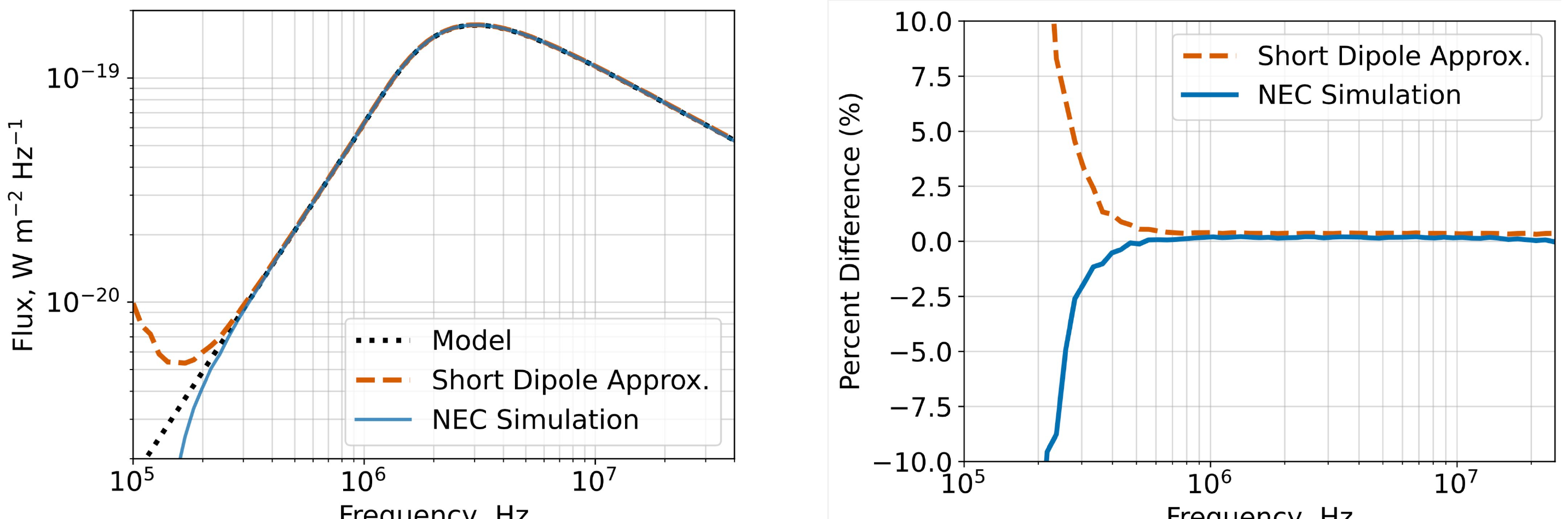}
        \caption{Left: The mean value of the reconstructed flux for the model (black dotted) the short dipole approximation (orange dashed) and NEC Simulation (blue) Right: The percent difference between the short dipole approximation (orange dashed) and NEC Simulation (blue) to the .}
    \label{fig:predicted_flux_comparisons}
\end{figure*}

 \begin{figure}[ht]
    \centering
    \includegraphics[width=.7\linewidth]{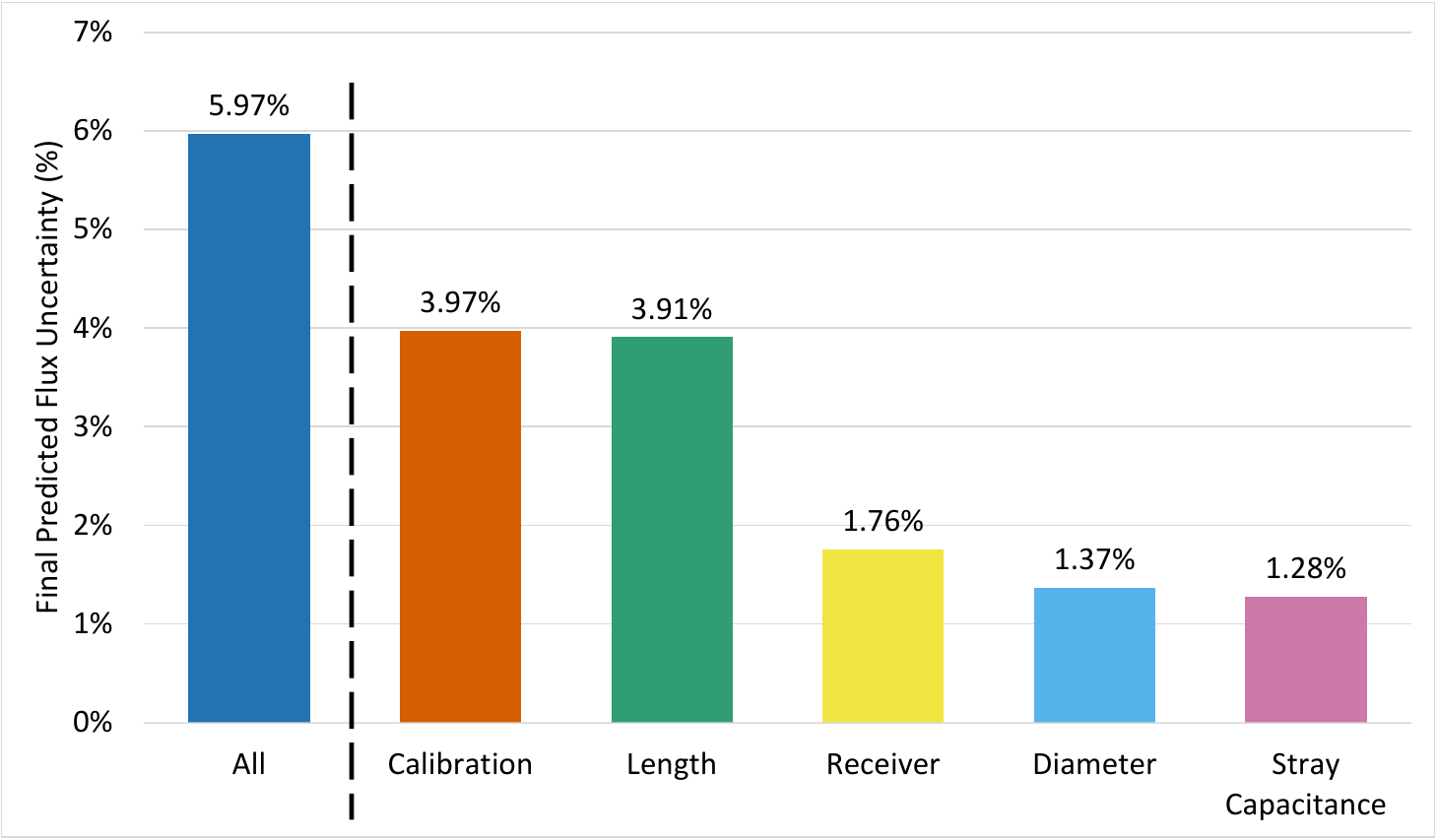}
    \caption{The predicted flux uncertainty at 10.5\,MHz for the component analysis. The blue bar gives the result using all uncertainty components. It matches the value of the NEC2 uncertainty in Figure~\ref{fig:flux_uncertainty}. The bars to the right of the dotted line give the predicted Galactic flux uncertainty when only uncertainty from the one labeled component is used. }
    \label{fig:single_variable_contributions}
\end{figure}

\section{Results}\label{sec:results}

Figure~\ref{fig:final_flux_uncertainties} shows the Measured Effective Flux obtained from simulated measurements $U'_m$ given by $S_{m}=U'_m/\chi$. The flux-equivalent contributions of simulated Plasma noise $S_p=U'_p/\chi$ and Amplifier noise $S_r=U'_r/\chi$ are also shown with their corresponding uncertainties as shaded band. The reconstructed Galactic flux $S_g=U_g'/\chi$ is shown as a light blue dashed line, with the corresponding measurement uncertainty shown as a band. The truth input is shown as a dotted blue line. The reconstructed Galactic flux used the Monte Carlo uncertainty propagation with the NEC2 antenna model. Each line gives the mean values of the posterior Monte Carlo distribution while the shaded regions represent the standard deviation. At frequencies below a few MHz, Figure~\ref{fig:final_flux_uncertainties} shows the reconstructed flux being less than the model, and the uncertainty increasing rapidly. This is due to the Plasma QTN (and associated uncertainty) dominating the calculation.

The estimated sky flux uncertainty is shown in Figure~\ref{fig:flux_uncertainty}. The solid blue line is obtained from the Monte Carlo method using the NEC2 antenna model, which is equivalent to the width of the blue colored band in Figure~\ref{fig:final_flux_uncertainties}. For comparison, the dashed orange line gives the uncertainty using the Monte Carlo method, but with using the electrically short dipole approximation instead of NEC2 simulations. The black dotted line gives the uncertainty using the analytical error propagation described in \ref{A:analytical}. The differences between the NEC2 model and the electrically short dipole approximation are consistent in the band of interest while the analytical estimate underestimates the uncertainty. We interpret this discrepancy as the result of the analytical model, while using some short dipole approximations, not including the covariance between various input parameters which the Monte Carlo simulation takes in by design. Figure~\ref{fig:predicted_flux_comparisons} shows the reconstructed sky flux using the Cane model, the MC with NEC simulation and the MC with short dipole approximation.

The impact of individual parameter uncertainty is not straightforward due to the nonlinear nature of some of the components in the model used. One way to investigate the impact of single components is to remove the uncertainty for every component except for one and repeat for each component in turn. The results of this analysis are shown in Figure~\ref{fig:single_variable_contributions}. Note these results are dependent on the uncertainty values chosen for each components.
This result highlights the dominant contributions from the calibration and length terms. In addition, stray capacitance monitoring is essential to reduce the relative contribution of its uncertainty (see \S\ref{sec:stray_cap_investigation}).

\section{Discussion}\label{sec:discussion}

The receiving system analyzed corresponds to a dipole mounted on a SmallSat, such as SunRISE~\cite{Kasper_2022}. We have allowed for parameters such as the antenna length and stray capacitance to vary. Antenna lengths and diameters can be chosen to suit the mission while minimizing the stray capacitance is a function of the space available at the location where the antennas are mounted, the geometry of its surrounding,  and contributions to the front-end electronics. 


\subsection{Importance of Plasma Uncertainty at Low Frequencies}
The magnitude of the uncertainty of the Galactic flux prediction is related to the relationship of the individual components as shown in Figure~\ref{fig:final_flux_uncertainties}. Above 1\,MHz, the predicted flux is dominated by measured flux uncertainty (mostly from the conversion from voltage to flux and the calibration noise) and the amplifier terms, with the plasma flux being negligible. Below 1\,MHz, the uncertainties from the plasma (and the amplifier to a lesser degree) become much closer to the final predicted results, significantly increasing the uncertainty. This result demonstrates that for measurements below 1\,MHz, the effect of the plasma noise becomes critically important and necessitates efforts to reduce the plasma uncertainty.

\subsection{Stray Capacitance Investigation}\label{sec:stray_cap_investigation}
\begin{figure}[ht]
    \centering
    \includegraphics[width=.65\linewidth]{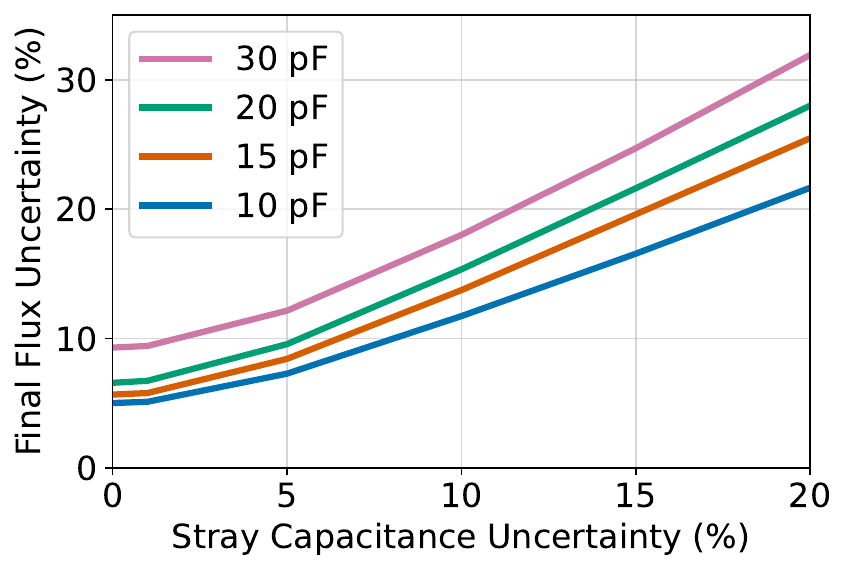}
    \caption{Predicted Galactic flux uncertainty versus the stray capacitance uncertainty, with four nominal values of stray capacitance.}
    \label{fig:stray_capacitance_investigation}
\end{figure}

In order to better understand the contribution of stray capacitance, the model was run using a range of stray capacitance values (10\,pF, 15\,pF, 20\,pf, and 30\,pf) and uncertainties (0\% to 20\%), with the results at 10.5\,MHz presented in Figure~\ref{fig:stray_capacitance_investigation}. All other uncertainties were included and remained constant. Both the stray capacitance and the uncertainty are proportional to the predicted flux uncertainty. Of note, reducing the stray capacitance uncertainty to 0\% can lead to a final flux uncertainty of between 9\% for 30\,pF and 5\% for 10\,pF. Examining the 15\,pF line demonstrates the importance of stray capacitance calibration systems discussed in \S\ref{sec:stray_capacitance}. At 1\% stray capacitance uncertainty, which is viable with a calibration system, the flux uncertainty is 5.97\%. Without a calibration system, an uncertainty of 10\% is more realistic, resulting in a flux uncertainty of 13\%. With a stray capacitance uncertainty of 10\%, the stray capacitance because the largest single contributor of flux uncertainty when repeating the analysis shown in Figure~\ref{fig:single_variable_contributions}.

\section{Conclusion}
This work presents an instrumental error budget for space-based measurements of the absolute flux of the sky synchrotron spectrum at frequencies below 20~MHz. The current understanding of the HF radio sky is limited and motivates further investigation. The results of this work demonstrate it is feasible to construct a spacecraft with low enough uncertainties to characterize Galactic synchrotron emission. Using NEC2 antenna simulations and Monte Carlo uncertainty propagation, the estimated Galactic flux uncertainty is approximately 6\% above 1 MHz, and increases rapidly at lower frequencies. This model could be used in adjusting design parameters in order to meet requirements.  As illustrated in Figure~\ref{fig:single_variable_contributions}, length and calibration uncertainty are the next largest contributors to the overall uncertainty on the reconstructed Galactic flux. In order to improve from our result, techniques for monitoring length should be developed and calibration parameters would need to be improved. Future work could involve probing uncertainties below 1\,MHz, particularly apropos minimizing statistical uncertainties and the large uncertainties associated with plasma quasi-thermal noise. In designing a spacecraft, consideration of stray capacitance is crucial, and necessitates a stray capacitance calibration and monitoring system to reduce overall uncertainties.

\section*{Data Availability Statement}
All results can be reproduced using data and software archived at \url{https://doi.org/10.5281/zenodo.11483379}~\cite{zenodo-software}. The PyGDSM library published at \url{https://ascl.net/1603.013}~\cite{pygdsm} and the Ultra Long wavelength Sky model with Absorption (ULSA) published at \url{https://doi.org/10.5281/zenodo.4663463}~\cite{ulsa_code} are required.

\appendix
\section{Finite Resistance and Reactance Equations}\label{A:finite dipole}
The NEC2 simulations and the electrically short dipole approximations of the radiation resistance and the antenna reactance is presented in Figure~\ref{fig:resistance_vs_length} and~\ref{fig:reactance_vs_length}. The disparity between approximation and the simulation can be partially resolved using analytical expressions for a finite dipole given in~\citeA{balanis-textbook}:

\begin{multline}
R_{rad} = \frac{\eta}{2\pi \sin^{2}(kL/2)} \biggl\{C + \ln{(kL)} - C_i(kL)
+ \frac{1}{2}\sin{(kL)} \left[ S_i(2kL) - 2S_i(kL)\right] \\
+ \frac{1}{2}\cos{(kL)} \left[ C + \ln{(kL/2)} + C_i(2kL) - 2C_i(kL)\right]\biggl\}
\end{multline}
And
\begin{multline}
X_a = \frac{\eta}{4\pi \sin^{2}(kL/2)} \biggl\{2S_i(kL) + \cos{(kL)}[2S_i(kL)-S_i(2kL)] \\
- \sin{(kL)} \left[ 2C_i(kL) - C_i(2kL) - C_i(\frac{2kr^2}{L})\right]\biggl\}
\end{multline}

where $\eta$ is the intrinsic impedance of the medium, $L$ is the full length, $k$ is the wave number, $C$ is Euler's constant, and $r$ is the radius. $C_i$ and $S_i$ are the cosine and sine integrals given by
\begin{equation}
C_i = -\int^{\infty}_{x} \frac{\cos y}{y} dy = \int^{x}_{\infty} \frac{\cos y}{y}
\end{equation}
And
\begin{equation}
S_i = \int^{x}_{0} \frac{\sin y}{y} dy
\end{equation}

Recreations of Figure~\ref{fig:resistance_vs_length} and~\ref{fig:reactance_vs_length} are given below, with the finite dipole equations used instead of the electrically short approximations.

\begin{figure*}[ht]
    \centering
    \includegraphics[width=1\linewidth]{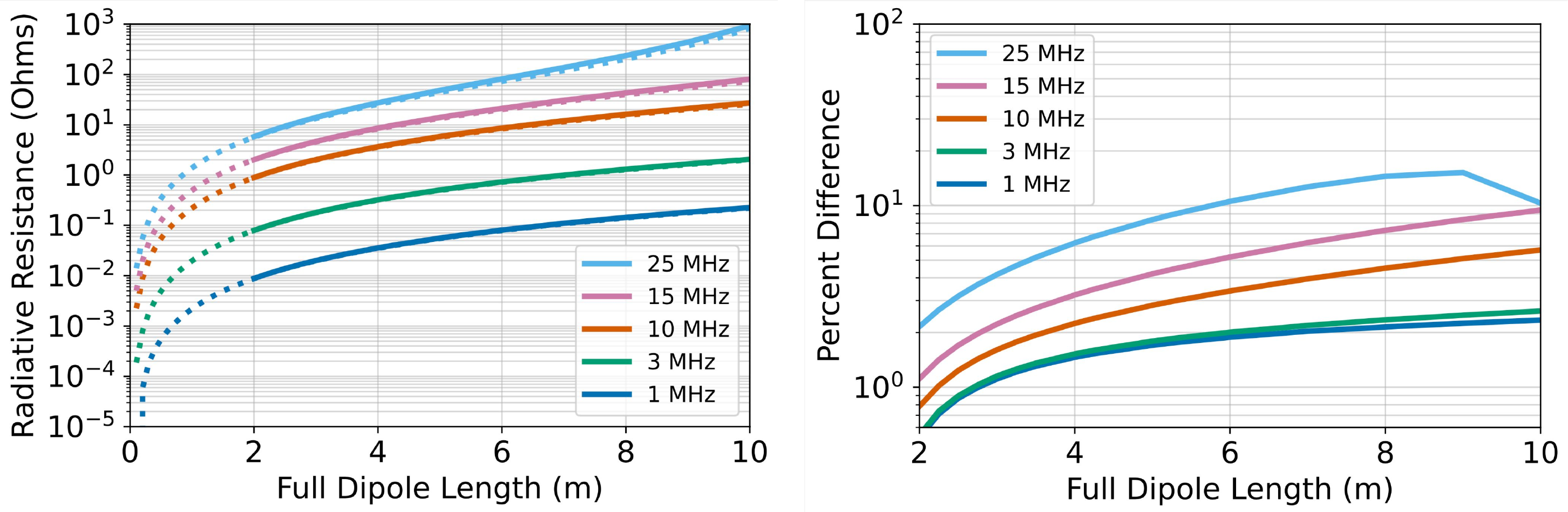}
        \caption{The finite equation version of Figure~\ref{fig:resistance_vs_length}. Left: The antenna radiative resistance versus full dipole length at five different frequencies. The NEC2 simulation shown by the solid line and the dotted line shows the finite dipole analytical expression. Right: The percent relative difference between the NEC2 simulation results and the analytical expression.}
    \label{fig:resistance_vs_length_exact}
\end{figure*}

\begin{figure*}[ht]
    \centering
    \includegraphics[width=1\linewidth]{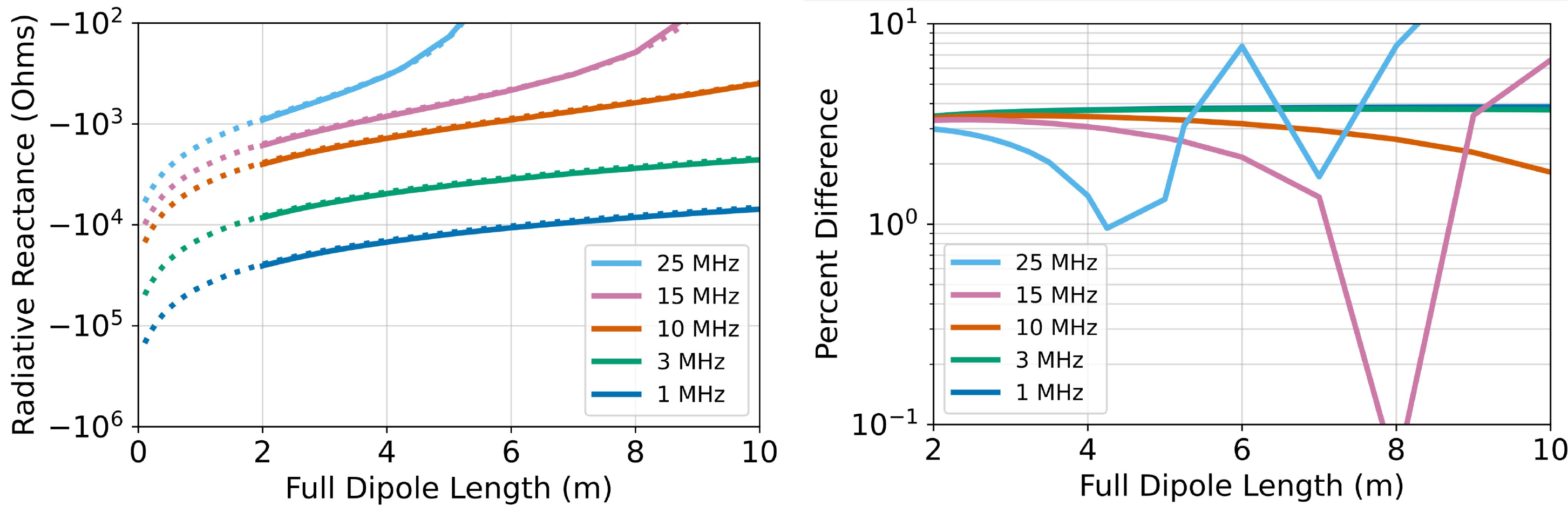}
        \caption{The finite equation version of Figure~\ref{fig:reactance_vs_length}. Left: The antenna reactance versus full dipole length at five different frequencies. The NEC2 simulation shown by the solid line and the dotted line shows the finite dipole analytical expression. Right: The percent absolute relative difference between the NEC2 simulation results and the analytical expression.}
    \label{fig:reactance_vs_length_exact}
\end{figure*}

\begin{figure*}[ht]
    \centering
    \includegraphics[width=1\linewidth]{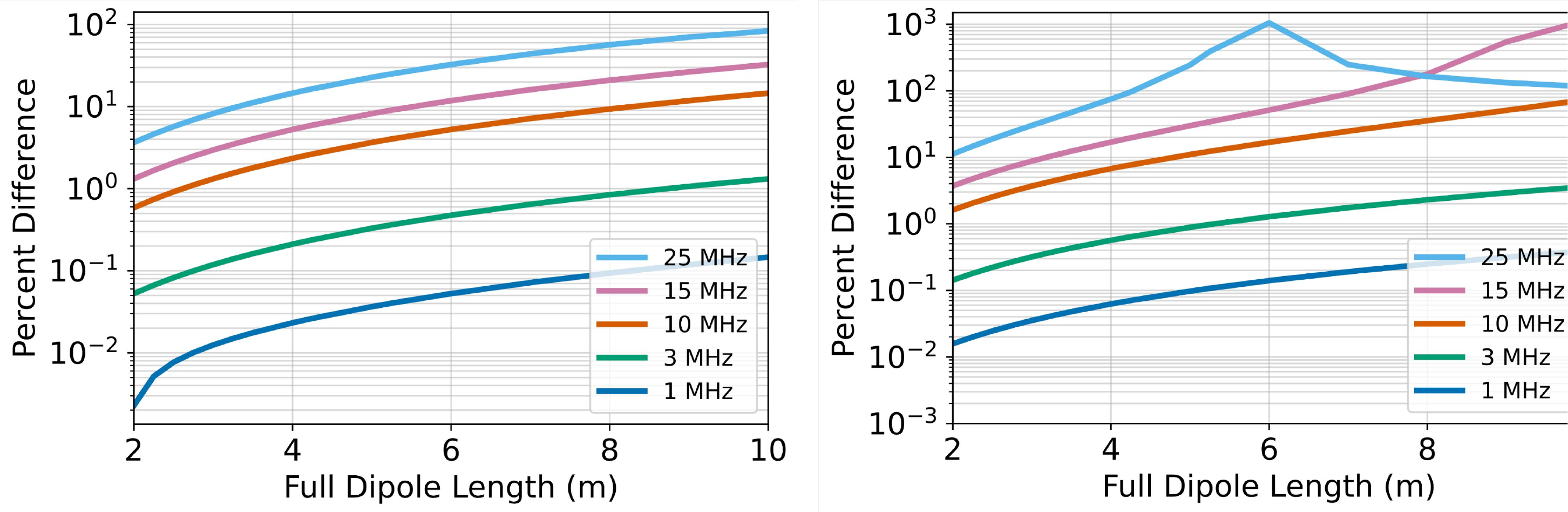}
        \caption{Percent difference between the electrically short dipole and the finite dipole equations for the radiative resistance (left) and the antenna reactance (right).}
    \label{fig:finite_compare_to_esd}
\end{figure*}

The percent difference between the short dipole approximation and the finite dipole equation is given in Figure~\ref{fig:finite_compare_to_esd}.

\section{Flux to Voltage Conversion Calculations} \label{A:flux to voltage}
This appendix describes the derivation of the flux to voltage conversion.  


The effective area, $A_e$, also called effective or equivalent aperture, is defined as the ratio of the available power at the terminals of a receiving antenna when a wave impinges on it
\begin{equation}\label{eq:Ae1}
A_e = \frac{P_T}{S} = \frac{|I_T|^2 R_t/2}{S}
\end{equation}
where $P_{T}$ is the power delivered to the load (in W) and $S$ is the power density of the incident wave (in W/m$^2$). The effective area generally depends on the direction of the incoming flux $\hat{\mathbf{r}}$ relative to the antenna and can be calculated in a number of different ways. 
In relation to the directivity, defined as the normalized intensity \break $D(\hat{\mathbf{r}})=4\pi U(\hat{\mathbf{r}})/\int_{4\pi} d\Omega \ U(\hat{\mathbf{r}})$,  the effective area is given by
\begin{equation}\label{eq:Ae2}
A_{e}(\hat{\mathbf{r}}) = \frac{\lambda^2}{4\pi} D(\hat{\mathbf{r}}).
\end{equation}
In terms of the impedance of free space, $Z_0$, the effective length, $\ell(\hat{\mathbf{r}})$, which is also dependent on the direction of the incident flux, and the radiative resistance, the effective area is given by
\begin{equation}\label{eq:Ae3}
A_e(\hat{\mathbf{r}}) = \frac{\ell^2(\hat{\mathbf{r}}) Z_0}{4 R_r}.
\end{equation}

Equating Eq.~\ref{eq:Ae2} and~\ref{eq:Ae3} and solving for $\ell^2$,
\begin{equation}\label{eq:ell}
\ell^2(\hat{\mathbf{r}}) = \frac{\lambda^2 D(\hat{\mathbf{r}})}{4\pi} \frac{4 R_r}{Z_0}.
\end{equation}


Consider, the voltage induced on the open circuit terminals when a wave impinges on it is related to the effective length, $\ell$,  and the voltage divider, $\Gamma$:
\begin{equation}\label{eq:voc}
V_{oc}(\hat{\mathbf{r}}) = E_i(\hat{\mathbf{r}}) \ell(\hat{\mathbf{r}}) \Gamma.
\end{equation}

The voltage squared produced by an $E_i(\hat{\mathbf{r}})$ field from an infinitesimal solid angle, $d\Omega$, in the direction $\hat{\mathbf{r}}$ is then given by 
\begin{equation}\label{eq:voc_squared}
d(V_{oc}^2(\hat{\mathbf{r}})) = E_i^2(\hat{\mathbf{r}}) \ell^2(\hat{\mathbf{r}}) \Gamma^2 d\Omega
\end{equation}

Substitute $E^2(\hat{\mathbf{r}})$ in terms of brightness ($E^2(\hat{\mathbf{r}}) = B(\hat{\mathbf{r}}) Z_0$) and the expression for effective length from Eq.~\ref{eq:ell}:
$$d(V_{oc}^2(\hat{\mathbf{r}})) = B(\hat{\mathbf{r}}) Z_0 \left(\frac{\lambda^2 D(\hat{\mathbf{r}})}{4\pi} \frac{4 R_r}{Z_0}\right) \Gamma^2 d\Omega$$

Integration of $d\Omega$ over the full $4\pi$ steriadan sky on both sides results in
\begin{equation}\label{eq:vos_squared_complete}
V_{oc}^2 = \left (\frac{R_r \lambda^2 }{\pi} \Gamma^2 \right ) \int_{4\pi} B(\hat{\mathbf{r}}) D(\hat{\mathbf{r}}) d\Omega
\end{equation}


The integral of the brightness over the antenna directivity gives the beam-averaged flux
\begin{equation}
\int_{4\pi} B(\hat{\mathbf{r}}) D(\hat{\mathbf{r}}) d\Omega = \langle S \rangle
\end{equation}



Substituting this result into Eq.~\ref{eq:vos_squared_complete}
\begin{equation}
V_{oc}^2 = \left (\frac{R_r \lambda^2 }{\pi} \Gamma^2 \right ) \langle S \rangle.
\end{equation}

The final flux to voltage conversion term comes from dividing both sides by the beam-averaged flux to give the flux-to-voltage conversion
\begin{equation}\label{eq:flux_to_voltage_conversion}
\chi \equiv \frac{V_{oc}^2}{\langle S \rangle} = \frac{R_r \lambda^2 }{\pi} \Gamma^2
\end{equation}

Note that while the $V_{oc}$ and  $\langle S \rangle$ both depend on the antenna beam pattern and the sky brightness distribution, the terms on the right do not. This relation assumes the antenna response is linear.

\section{Analytical Uncertainty Propagation Calculations}\label{A:analytical}
Analytical uncertainty propagation was done for comparison to the Monte Carlo technique. This involved describing the uncertainty of the parameters using partial derivatives and combining appropriately. This appendix describes the formalism and steps to achieve the analytical uncertainty result.

The antenna capacitance is given by:
\begin{equation}
C_{a}=\frac{2 \pi \epsilon_{0} L}{\ln \left(\frac{L}{a}\right)-1}
\end{equation}
The uncertainty is given by:
\begin{equation}
\Delta C_{a}^{2}=\Delta L^{2}\left|\frac{\partial C_{a}}{\partial L}\right|^{2}+\Delta a^{2}\left|\frac{\partial C_{a}}{\partial a}\right|^{2}
\end{equation}
Taking the derivative with respect to $L$ and $d$ :
\begin{equation}
\begin{gathered}
\frac{\partial C_{a}}{\partial L}=\frac{2 \pi \epsilon_{0}\left[\ln \left(\frac{L}{a}\right)-2\right]}{[\ln \left(\frac{L}{a}\right)-1]^{2}} \\
\frac{\partial C_{a}}{\partial a}=\frac{2 \pi \epsilon_{0} L}{a\left[\ln \left(\frac{L}{a}\right)-1\right]^{2}}
\end{gathered}
\end{equation}
Then
\begin{equation}
\Delta C_{a}^{2}=\left(\Delta L \frac{2 \pi \epsilon_{0}\left[\ln \left(\frac{L}{a}\right)-2\right]}{\left[\ln \left(\frac{L}{a}\right)-1\right]^{2}}\right)^{2}+\left(\Delta a \frac{2 \pi \epsilon_{0} L}{a\left[\ln \left(\frac{L}{a}\right)-1\right]^{2}}\right)^{2}
\end{equation}
Dividing by the expression for $C_{a}$ gives the relative uncertainty:
\begin{equation}
\begin{gathered}
\left(\frac{\Delta C_{a}}{C_{a}}\right)^{2}=\left(\Delta L \frac{\left[\ln \left(\frac{L}{a}\right)-2\right]}{L\left[\ln \left(\frac{L}{a}\right)-1\right]}\right)^{2}+\left(\Delta a \frac{1}{a\left[\ln \left(\frac{L}{a}\right)-1\right]}\right)^{2} \\
\left(\frac{\Delta C_{a}}{C_{a}}\right)^{2}=\left(\frac{\Delta L}{L} \frac{\ln \left(\frac{L}{a}\right)-2}{\ln \left(\frac{L}{a}\right)-1}\right)^{2}+\left(\frac{\Delta a}{a} \frac{1}{\ln \left(\frac{L}{a}\right)-1}\right)^{2}
\end{gathered}
\end{equation}

\noindent In the limit of small antenna resistance, and at low frequencies where $X\simeq -1/(2\pi f C )$, a simplified version of the voltage divider is given by:
\begin{equation}
\Gamma=\left(1+\frac{C_{s}}{C_{a}}\right)^{-2}
\end{equation}
The uncertainty is given by:
\begin{equation}
\Delta \Gamma^{2}=\Delta C_{s}^{2}\left|\frac{\partial \Gamma}{\partial C_{s}}\right|^{2}+\Delta C_{a}^{2}\left|\frac{\partial \Gamma}{\partial C_{a}}\right|^{2}
\end{equation}
Taking the derivative with respect to $C_{s}$ and $C_{a}$ :
\begin{equation}
\begin{gathered}
\frac{\partial \Gamma}{\partial C_{s}}=-\frac{2}{C_{a}}\left(1+\frac{C_{s}}{C_{a}}\right)^{-3} \\
\frac{\partial \Gamma}{\partial C_{a}}=-2 \frac{C_{s}}{C_{a}}\left(1+\frac{C_{S}}{C_{a}}\right)^{-3}
\end{gathered}
\end{equation}
Then
\begin{equation}
\Delta \Gamma^{2}=\left(\Delta C_{s} \frac{2}{C_{a}}\left(1+\frac{C_{s}}{C_{a}}\right)^{-3}\right)^{2}+\left(\Delta C_{a} 2 \frac{C_{s}}{C_{a}^{2}}\left(1+\frac{C_{s}}{C_{a}}\right)^{-3}\right)^{2}
\end{equation}
Dividing by the expression for $\Gamma$ gives the relative uncertainty:
\begin{equation}
\begin{aligned}
&\left(\frac{\Delta \Gamma}{\Gamma}\right)^{2}=\left(\Delta C_{s} \frac{2}{C_{a}}\left(1+\frac{C_{s}}{C_{a}}\right)^{-1}\right)^{2}+\left(\Delta C_{a} 2 \frac{C_{s}}{C_{a}^{2}}\left(1+\frac{C_{s}}{C_{a}}\right)^{-1}\right)^{2} \\
&\left(\frac{\Delta \Gamma}{\Gamma}\right)^{2}=\left(\frac{\Delta C_{s}}{C_{s}} 2 \frac{C_{s}}{C_{a}}\left(1+\frac{C_{s}}{C_{a}}\right)^{-1}\right)^{2}+\left(\frac{\Delta C_{a}}{C_{a}} 2 \frac{C_{s}}{C_{a}}\left(1+\frac{C_{s}}{C_{a}}\right)^{-1}\right)^{2}
\end{aligned}
\end{equation}

\noindent The voltage to flux term is given by:
\begin{equation}
\chi=x \frac{L_{m}}{\left(1+\frac{C_{S}}{C_{a}}\right)^{2}}
\end{equation}
Where $x$ is the constant $80\pi\,\Omega$.
The uncertainty is given by:
\begin{equation}
\Delta \chi^{2}=\Delta L_{m}^{2}\left|\frac{\partial \chi}{\partial L_{m}}\right|^{2}+\Delta C_{s}^{2}\left|\frac{\partial \chi}{\partial C_{s}}\right|^{2}+\Delta C_{a}^{2}\left|\frac{\partial \chi}{\partial C_{a}}\right|^{2}
\end{equation}
Taking the derivative with respect to $L_{m}, C_{s}$ and $C_{a}$ :
\begin{equation}
\begin{aligned}
\frac{\partial \chi}{\partial L} &=\frac{x}{\left(1+\frac{C_{s}}{C_{a}}\right)^{2}} \\
\frac{\partial \chi}{\partial C_{s}} &=\frac{2 C_{a}^{2} L x}{\left(C_{a}+C_{s}\right)^{3}} \\
\frac{\partial \chi}{\partial C_{a}} &=\frac{2 C_{a} C_{s} L x}{\left(C_{a}+C_{s}\right)^{3}}
\end{aligned}
\end{equation}
Then
\begin{equation}
\Delta \chi^{2}=\left(\Delta L_{m} \frac{x}{\left(1+\frac{C_{s}}{C_{a}}\right)^{2}}\right)^{2}+\left(\Delta C_{s} \frac{2 L x}{C_{a}\left(1+\frac{C_{S}}{C_{a}}\right)^{3}}\right)^{2}+\left(\Delta C_{a} \frac{2 C_{s} L x}{C_{a}^{2}\left(1+\frac{C_{s}}{C_{a}}\right)^{3}}\right)^{2}
\end{equation}
Dividing by the expression for $\chi$ gives the relative uncertainty:
\begin{equation}
\begin{gathered}
\left(\frac{\Delta \chi}{\chi}\right)^{2}=\Delta L_{m}^{2}+\left(\Delta C_{s} \frac{2\left(1+\frac{C_{s}}{C_{a}}\right)^{2}}{C_{a}\left(1+\frac{C_{s}}{C_{a}}\right)^{3}}\right)^{2}+\left(\Delta C_{a} \frac{2 C_{s}\left(1+\frac{C_{s}}{C_{a}}\right)^{2}}{C_{a}^{2}\left(1+\frac{C_{s}}{C_{a}}\right)^{3}}\right)^{2} \end{gathered}
\end{equation}
\begin{equation}\label{eq.conversion_uncertainty}
\begin{gathered}
\left(\frac{\Delta \chi}{\chi}\right)^{2}=\left(L_{m} \frac{\Delta L_{m}}{L_{m}}\right)^{2}+\left(\frac{\Delta C_{s}}{C_{s}} \frac{C_{s}}{C_{a}} \frac{2}{\left(1+\frac{C_{s}}{C_{a}}\right)}\right)^{2}+\left(\frac{\Delta C_{a}}{C_{a}} \frac{C_{s}}{C_{a}} \frac{2}{\left(1+\frac{C_{s}}{C_{a}}\right)}\right)^{2}
\end{gathered}
\end{equation}

The plasma voltage at the antenna is given by:
\begin{equation}
U_{p}=5 \times 10^{-5} \frac{n_{e} T_{e}}{f^{3} L_{m}} \Gamma
\end{equation}
Since there are only multiplicative terms in the expression for $U_{p}$, the relative uncertainty is just the relative uncertainty of the components added in quadrature:
\begin{equation}
\left(\frac{\Delta U_{p}}{U_{p}}\right)^{2}=\left(\frac{\Delta L_{m}}{L_{m}}\right)^{2}+\left(\frac{\Delta n_{e}}{n_{e}}\right)^{2}+\left(\frac{\Delta T_{e}}{T_{e}}\right)^{2}+\left(\frac{\Delta \Gamma}{\Gamma}\right)^{2}
\end{equation}

The voltage and uncertainty from the amplifier is given by 
\begin{equation}
U_a = U_{V} + U_{I} + +U_{J}
\end{equation}
\begin{equation}
\Delta U_{a}=\sqrt{\Delta U_{V}^{2}+\Delta U_{I}^{2}+\Delta U_{j}^{2}}
\end{equation}

The operational amplifier voltage noise, $U_V$, is simply $V_n^2$, where $V_n$ is the voltage noise density specification of the component used. The operational amplifier current noise, $U_{I}$, is $(I_n\times |Z_{eq}|)^2$, where $I_n$ is the current noise density specification of the component used, and $Z_{eq}$ is the impedance of the circuit looking out from the input terminals of the amplifier. Finally, the Johnson–Nyquist thermal noise is given by $U_{J} = 4k_B T_0 R_T$, where $k_B$ is the Boltzmann constant, $T_0$ is the temperature of the system, and $R_T$ is the total resistance.

The received voltage from the galactic flux is given by:
\begin{equation}
U_{g}=U_{m}-U_{p}-U_{a}
\end{equation}
The effective Galactic voltage uncertainty is thus given by
\begin{equation}
\Delta U_{g}=\sqrt{\Delta U_{m}^{2}+\Delta U_{p}^{2}+\Delta U_{a}^{2}}
\end{equation}
We need to use absolute uncertainties, because although the relative uncertainties for each of these expressions are constant, the actual values vary with frequency.
\begin{equation}\label{eg.ug_uncertainty}
\Delta U_{g}^{2}=\left(\frac{\Delta U_{m}}{U_{m}} U_{m}\right)^{2}+\left(\frac{\Delta U_{p}}{U_{p}} U_{p}\right)^{2}+\left(\frac{\Delta U_{a}}{U_{a}} U_{a}\right)^{2}
\end{equation}
where $U_M$ is described simply by the calibration components of uncertainty added in quadrature.
Finally, the predicted flux is given by
\begin{equation}
S_{g}=\frac{U_{g}}{\chi}
\end{equation}
The relative uncertainty is just the relative uncertainty of the components added in quadrature, which have already been derived above in Equations~\ref{eg.ug_uncertainty} and~\ref{eq.conversion_uncertainty}:
\begin{equation}
\begin{gathered}
\frac{\Delta S_{g}}{S_{g}}=\sqrt{\left(\frac{\Delta U_{g}}{U_{g}}\right)^{2}+\left(\frac{\Delta \chi}{\chi}\right)^{2}} \end{gathered}
\end{equation}


\acknowledgments
This research has made use of the Astrophysics Data System, funded by NASA under Cooperative Agreement 80NSSC21M00561.
The research was carried out at the Jet Propulsion Laboratory, California Institute of Technology, under a contract with the National Aeronautics and Space Administration (80NM0018D0004) and funded through the President’s and Director’s Research and Development Fund (PDRDF).






\bibliography{main}

\begin{thebibliography}{}

\bibitem [\protect \citeauthoryear {%
Aghanim%
\ \protect \BOthers {.}}{%
Aghanim%
\ \protect \BOthers {.}}{%
{\protect \APACyear {2020}}%
}]{%
Planck-2020}
\APACinsertmetastar {%
Planck-2020}%
\begin{APACrefauthors}%
Aghanim, N.%
, Akrami, Y.%
, Arroja, F.%
, Ashdown, M.%
, Aumont, J.%
, Baccigalupi, C.%
\BDBL {}others%
\end{APACrefauthors}%
\unskip\
\newblock
\APACrefYearMonthDay{2020}{}{}.
\newblock
{\BBOQ}\APACrefatitle {Planck 2018 results-I. Overview and the cosmological legacy of Planck} {Planck 2018 results-i. overview and the cosmological legacy of planck}.{\BBCQ}
\newblock
\APACjournalVolNumPages{Astronomy \& Astrophysics}{641}{}{A1}.
\PrintBackRefs{\CurrentBib}

\bibitem [\protect \citeauthoryear {%
{Alexander}%
}{%
{Alexander}%
}{%
{\protect \APACyear {1970}}%
}]{%
1970PASP...82R1026A}
\APACinsertmetastar {%
1970PASP...82R1026A}%
\begin{APACrefauthors}%
{Alexander}, J\BPBI K.%
\end{APACrefauthors}%
\unskip\
\newblock
\APACrefYearMonthDay{1970}{{\APACmonth{01}}}{}.
\newblock
{\BBOQ}\APACrefatitle {{Galactic continuum radiation observations below 6 MHz from the Radio Astronomy Explorer satellite.}} {{Galactic continuum radiation observations below 6 MHz from the Radio Astronomy Explorer satellite.}}{\BBCQ}
\newblock
\APACjournalVolNumPages{"Publ.\ Astron.\ Pacific"}{82}{}{1026}.
\PrintBackRefs{\CurrentBib}

\bibitem [\protect \citeauthoryear {%
{Alexander}%
, {Brown}%
, {Clark}%
, {Stone}%
\BCBL {}\ \BBA {} {Weber}%
}{%
{Alexander}%
\ \protect \BOthers {.}}{%
{\protect \APACyear {1969}}%
}]{%
Alexander_1969}
\APACinsertmetastar {%
Alexander_1969}%
\begin{APACrefauthors}%
{Alexander}, J\BPBI K.%
, {Brown}, L\BPBI W.%
, {Clark}, T\BPBI A.%
, {Stone}, R\BPBI G.%
\BCBL {}\ \BBA {} {Weber}, R\BPBI R.%
\end{APACrefauthors}%
\unskip\
\newblock
\APACrefYearMonthDay{1969}{{\APACmonth{09}}}{}.
\newblock
{\BBOQ}\APACrefatitle {{The Spectrum of the Cosmic Radio Background Between 0.4 and 6.5 MHz}} {{The Spectrum of the Cosmic Radio Background Between 0.4 and 6.5 MHz}}.{\BBCQ}
\newblock
\APACjournalVolNumPages{The Astrophysical Journal Letters}{157}{}{L163}.
\newblock
\begin{APACrefDOI} \doi{10.1086/180411} \end{APACrefDOI}
\PrintBackRefs{\CurrentBib}

\bibitem [\protect \citeauthoryear {%
{Alexander}%
, {Kaiser}%
, {Novaco}%
, {Grena}%
\BCBL {}\ \BBA {} {Weber}%
}{%
{Alexander}%
\ \protect \BOthers {.}}{%
{\protect \APACyear {1975}}%
}]{%
1975A&A....40..365A}
\APACinsertmetastar {%
1975A&A....40..365A}%
\begin{APACrefauthors}%
{Alexander}, J\BPBI K.%
, {Kaiser}, M\BPBI L.%
, {Novaco}, J\BPBI C.%
, {Grena}, F\BPBI R.%
\BCBL {}\ \BBA {} {Weber}, R\BPBI R.%
\end{APACrefauthors}%
\unskip\
\newblock
\APACrefYearMonthDay{1975}{{\APACmonth{05}}}{}.
\newblock
{\BBOQ}\APACrefatitle {{Scientific instrumentation of the Radio-Astronomy-Explorer-2 satellite.}} {{Scientific instrumentation of the Radio-Astronomy-Explorer-2 satellite.}}{\BBCQ}
\newblock
\APACjournalVolNumPages{aap}{40}{4}{365-371}.
\PrintBackRefs{\CurrentBib}

\bibitem [\protect \citeauthoryear {%
Analog~Devices%
}{%
Analog~Devices%
}{%
{\protect \APACyear {2009}}%
}]{%
AD_app_note}
\APACinsertmetastar {%
AD_app_note}%
\begin{APACrefauthors}%
Analog~Devices, I.%
\end{APACrefauthors}%
\unskip\
\newblock
\APACrefYearMonthDay{2009}{}{}.
\newblock
{\BBOQ}\APACrefatitle {Op Amp Total Output Noise Calculations for Single-Pole System} {Op amp total output noise calculations for single-pole system}.{\BBCQ}
\newblock
\APACjournalVolNumPages{Analog Devices Tutorial}{MT-049}{}{}.
\PrintBackRefs{\CurrentBib}

\bibitem [\protect \citeauthoryear {%
Balanis%
}{%
Balanis%
}{%
{\protect \APACyear {2016}}%
}]{%
balanis-textbook}
\APACinsertmetastar {%
balanis-textbook}%
\begin{APACrefauthors}%
Balanis, C\BPBI A.%
\end{APACrefauthors}%
\unskip\
\newblock
\APACrefYear{2016}.
\newblock
\APACrefbtitle {Antenna theory: analysis and design} {Antenna theory: analysis and design}.
\newblock
\APACaddressPublisher{}{John Wiley \& sons}.
\PrintBackRefs{\CurrentBib}

\bibitem [\protect \citeauthoryear {%
Bale%
\ \protect \BOthers {.}}{%
Bale%
\ \protect \BOthers {.}}{%
{\protect \APACyear {2023}}%
}]{%
LuSEE2023}
\APACinsertmetastar {%
LuSEE2023}%
\begin{APACrefauthors}%
Bale, S\BPBI D.%
, Bassett, N.%
, Burns, J\BPBI O.%
, Jones, J\BPBI D.%
, Goetz, K.%
, Hellum-Bye, C.%
\BDBL {}Suzuki, A.%
\end{APACrefauthors}%
\unskip\
\newblock
\APACrefYearMonthDay{2023}{}{}.
\newblock
\APACrefbtitle {LuSEE 'Night': The Lunar Surface Electromagnetics Experiment.} {Lusee 'night': The lunar surface electromagnetics experiment.}
\PrintBackRefs{\CurrentBib}

\bibitem [\protect \citeauthoryear {%
{Bale}%
\ \protect \BOthers {.}}{%
{Bale}%
\ \protect \BOthers {.}}{%
{\protect \APACyear {2016}}%
}]{%
2016SSRv..204...49B}
\APACinsertmetastar {%
2016SSRv..204...49B}%
\begin{APACrefauthors}%
{Bale}, S\BPBI D.%
, {Goetz}, K.%
, {Harvey}, P\BPBI R.%
, {Turin}, P.%
, {Bonnell}, J\BPBI W.%
, {Dudok de Wit}, T.%
\BDBL {}{Wygant}, J\BPBI R.%
\end{APACrefauthors}%
\unskip\
\newblock
\APACrefYearMonthDay{2016}{{\APACmonth{12}}}{}.
\newblock
{\BBOQ}\APACrefatitle {{The FIELDS Instrument Suite for Solar Probe Plus. Measuring the Coronal Plasma and Magnetic Field, Plasma Waves and Turbulence, and Radio Signatures of Solar Transients}} {{The FIELDS Instrument Suite for Solar Probe Plus. Measuring the Coronal Plasma and Magnetic Field, Plasma Waves and Turbulence, and Radio Signatures of Solar Transients}}.{\BBCQ}
\newblock
\APACjournalVolNumPages{ssr}{204}{1-4}{49-82}.
\newblock
\begin{APACrefDOI} \doi{10.1007/s11214-016-0244-5} \end{APACrefDOI}
\PrintBackRefs{\CurrentBib}

\bibitem [\protect \citeauthoryear {%
Bale%
\ \protect \BOthers {.}}{%
Bale%
\ \protect \BOthers {.}}{%
{\protect \APACyear {2008}}%
}]{%
StereoWaves-antennas}
\APACinsertmetastar {%
StereoWaves-antennas}%
\begin{APACrefauthors}%
Bale, S\BPBI D.%
, Ullrich, R.%
, Goetz, K.%
, Alster, N.%
, Cecconi, B.%
, Dekkali, M.%
\BDBL {}others%
\end{APACrefauthors}%
\unskip\
\newblock
\APACrefYearMonthDay{2008}{}{}.
\newblock
{\BBOQ}\APACrefatitle {The electric antennas for the STEREO/WAVES experiment} {The electric antennas for the stereo/waves experiment}.{\BBCQ}
\newblock
\APACjournalVolNumPages{The STEREO Mission}{}{}{529--547}.
\PrintBackRefs{\CurrentBib}

\bibitem [\protect \citeauthoryear {%
{Bale}%
\ \protect \BOthers {.}}{%
{Bale}%
\ \protect \BOthers {.}}{%
{\protect \APACyear {2008}}%
}]{%
2008SSRv..136..529B}
\APACinsertmetastar {%
2008SSRv..136..529B}%
\begin{APACrefauthors}%
{Bale}, S\BPBI D.%
, {Ullrich}, R.%
, {Goetz}, K.%
, {Alster}, N.%
, {Cecconi}, B.%
, {Dekkali}, M.%
\BDBL {}{Pulupa}, M.%
\end{APACrefauthors}%
\unskip\
\newblock
\APACrefYearMonthDay{2008}{{\APACmonth{04}}}{}.
\newblock
{\BBOQ}\APACrefatitle {{The Electric Antennas for the STEREO/WAVES Experiment}} {{The Electric Antennas for the STEREO/WAVES Experiment}}.{\BBCQ}
\newblock
\APACjournalVolNumPages{ssr}{136}{1-4}{529-547}.
\newblock
\begin{APACrefDOI} \doi{10.1007/s11214-007-9251-x} \end{APACrefDOI}
\PrintBackRefs{\CurrentBib}

\bibitem [\protect \citeauthoryear {%
{Bassett}%
\ \protect \BOthers {.}}{%
{Bassett}%
\ \protect \BOthers {.}}{%
{\protect \APACyear {2023}}%
}]{%
Bassett_2023}
\APACinsertmetastar {%
Bassett_2023}%
\begin{APACrefauthors}%
{Bassett}, N.%
, {Rapetti}, D.%
, {Nhan}, B\BPBI D.%
, {Page}, B.%
, {Burns}, J\BPBI O.%
, {Pulupa}, M.%
\BCBL {}\ \BBA {} {Bale}, S\BPBI D.%
\end{APACrefauthors}%
\unskip\
\newblock
\APACrefYearMonthDay{2023}{{\APACmonth{03}}}{}.
\newblock
{\BBOQ}\APACrefatitle {{Constraining a Model of the Radio Sky below 6 MHz Using the Parker Solar Probe/FIELDS Instrument in Preparation for Upcoming Lunar-based Experiments}} {{Constraining a Model of the Radio Sky below 6 MHz Using the Parker Solar Probe/FIELDS Instrument in Preparation for Upcoming Lunar-based Experiments}}.{\BBCQ}
\newblock
\APACjournalVolNumPages{The Astrophysical Journal}{945}{2}{134}.
\newblock
\begin{APACrefDOI} \doi{10.3847/1538-4357/acbc76} \end{APACrefDOI}
\PrintBackRefs{\CurrentBib}

\bibitem [\protect \citeauthoryear {%
Baumjohann%
\ \BBA {} Treumann%
}{%
Baumjohann%
\ \BBA {} Treumann%
}{%
{\protect \APACyear {1996}}%
}]{%
plasma_properties}
\APACinsertmetastar {%
plasma_properties}%
\begin{APACrefauthors}%
Baumjohann, W.%
\BCBT {}\ \BBA {} Treumann, R\BPBI A.%
\end{APACrefauthors}%
\unskip\
\newblock
\APACrefYear{1996}.
\newblock
\APACrefbtitle {Basic Space Plasma Physics} {Basic space plasma physics}.
\newblock
\APACaddressPublisher{}{Imperial College Press}.
\newblock
\begin{APACrefDOI} \doi{10.1142/p015} \end{APACrefDOI}
\PrintBackRefs{\CurrentBib}

\bibitem [\protect \citeauthoryear {%
Bennett%
\ \protect \BOthers {.}}{%
Bennett%
\ \protect \BOthers {.}}{%
{\protect \APACyear {2013}}%
}]{%
WMAP-2013}
\APACinsertmetastar {%
WMAP-2013}%
\begin{APACrefauthors}%
Bennett, C\BPBI L.%
, Larson, D.%
, Weiland, J\BPBI L.%
, Jarosik, N.%
, Hinshaw, G.%
, Odegard, N.%
\BDBL {}others%
\end{APACrefauthors}%
\unskip\
\newblock
\APACrefYearMonthDay{2013}{}{}.
\newblock
{\BBOQ}\APACrefatitle {Nine-year Wilkinson Microwave Anisotropy Probe (WMAP) observations: final maps and results} {Nine-year wilkinson microwave anisotropy probe (wmap) observations: final maps and results}.{\BBCQ}
\newblock
\APACjournalVolNumPages{The Astrophysical Journal Supplement Series}{208}{2}{20}.
\PrintBackRefs{\CurrentBib}

\bibitem [\protect \citeauthoryear {%
{Blumenthal}%
\ \BBA {} {Gould}%
}{%
{Blumenthal}%
\ \BBA {} {Gould}%
}{%
{\protect \APACyear {1970}}%
}]{%
1970RvMP...42..237B}
\APACinsertmetastar {%
1970RvMP...42..237B}%
\begin{APACrefauthors}%
{Blumenthal}, G\BPBI R.%
\BCBT {}\ \BBA {} {Gould}, R\BPBI J.%
\end{APACrefauthors}%
\unskip\
\newblock
\APACrefYearMonthDay{1970}{{\APACmonth{01}}}{}.
\newblock
{\BBOQ}\APACrefatitle {{Bremsstrahlung, Synchrotron Radiation, and Compton Scattering of High-Energy Electrons Traversing Dilute Gases}} {{Bremsstrahlung, Synchrotron Radiation, and Compton Scattering of High-Energy Electrons Traversing Dilute Gases}}.{\BBCQ}
\newblock
\APACjournalVolNumPages{Reviews of Modern Physics}{42}{2}{237-271}.
\newblock
\begin{APACrefDOI} \doi{10.1103/RevModPhys.42.237} \end{APACrefDOI}
\PrintBackRefs{\CurrentBib}

\bibitem [\protect \citeauthoryear {%
{Bougeret}%
\ \protect \BOthers {.}}{%
{Bougeret}%
\ \protect \BOthers {.}}{%
{\protect \APACyear {2008}}%
}]{%
2008SSRv..136..487B}
\APACinsertmetastar {%
2008SSRv..136..487B}%
\begin{APACrefauthors}%
{Bougeret}, J\BPBI L.%
, {Goetz}, K.%
, {Kaiser}, M\BPBI L.%
, {Bale}, S\BPBI D.%
, {Kellogg}, P\BPBI J.%
, {Maksimovic}, M.%
\BDBL {}{Zouganelis}, I.%
\end{APACrefauthors}%
\unskip\
\newblock
\APACrefYearMonthDay{2008}{{\APACmonth{04}}}{}.
\newblock
{\BBOQ}\APACrefatitle {{S/WAVES: The Radio and Plasma Wave Investigation on the STEREO Mission}} {{S/WAVES: The Radio and Plasma Wave Investigation on the STEREO Mission}}.{\BBCQ}
\newblock
\APACjournalVolNumPages{ssr}{136}{1-4}{487-528}.
\newblock
\begin{APACrefDOI} \doi{10.1007/s11214-007-9298-8} \end{APACrefDOI}
\PrintBackRefs{\CurrentBib}

\bibitem [\protect \citeauthoryear {%
{Bougeret}%
\ \protect \BOthers {.}}{%
{Bougeret}%
\ \protect \BOthers {.}}{%
{\protect \APACyear {1995}}%
}]{%
1995SSRv...71..231B}
\APACinsertmetastar {%
1995SSRv...71..231B}%
\begin{APACrefauthors}%
{Bougeret}, J\BPBI L.%
, {Kaiser}, M\BPBI L.%
, {Kellogg}, P\BPBI J.%
, {Manning}, R.%
, {Goetz}, K.%
, {Monson}, S\BPBI J.%
\BDBL {}{Hoang}, S.%
\end{APACrefauthors}%
\unskip\
\newblock
\APACrefYearMonthDay{1995}{{\APACmonth{02}}}{}.
\newblock
{\BBOQ}\APACrefatitle {{Waves: The Radio and Plasma Wave Investigation on the Wind Spacecraft}} {{Waves: The Radio and Plasma Wave Investigation on the Wind Spacecraft}}.{\BBCQ}
\newblock
\APACjournalVolNumPages{ssr}{71}{1-4}{231-263}.
\newblock
\begin{APACrefDOI} \doi{10.1007/BF00751331} \end{APACrefDOI}
\PrintBackRefs{\CurrentBib}

\bibitem [\protect \citeauthoryear {%
{Brown}%
}{%
{Brown}%
}{%
{\protect \APACyear {1973}}%
{\protect \APACexlab {{\protect \BCnt {1}}}}}]{%
1973ApJ...180..359B}
\APACinsertmetastar {%
1973ApJ...180..359B}%
\begin{APACrefauthors}%
{Brown}, L\BPBI W.%
\end{APACrefauthors}%
\unskip\
\newblock
\APACrefYearMonthDay{1973{\protect \BCnt {1}}}{{\APACmonth{03}}}{}.
\newblock
{\BBOQ}\APACrefatitle {{The Galactic Radio Spectrum Between 130 and 26OOkHz}} {{The Galactic Radio Spectrum Between 130 and 26OOkHz}}.{\BBCQ}
\newblock
\APACjournalVolNumPages{The Astrophysical Journal}{180}{}{359-370}.
\newblock
\begin{APACrefDOI} \doi{10.1086/151968} \end{APACrefDOI}
\PrintBackRefs{\CurrentBib}

\bibitem [\protect \citeauthoryear {%
{Brown}%
}{%
{Brown}%
}{%
{\protect \APACyear {1973}}%
{\protect \APACexlab {{\protect \BCnt {2}}}}}]{%
Brown_1973}
\APACinsertmetastar {%
Brown_1973}%
\begin{APACrefauthors}%
{Brown}, L\BPBI W.%
\end{APACrefauthors}%
\unskip\
\newblock
\APACrefYearMonthDay{1973{\protect \BCnt {2}}}{{\APACmonth{03}}}{}.
\newblock
{\BBOQ}\APACrefatitle {{The Galactic Radio Spectrum Between 130 and 26OOkHz}} {{The Galactic Radio Spectrum Between 130 and 26OOkHz}}.{\BBCQ}
\newblock
\APACjournalVolNumPages{The Astrophysical Journal}{180}{}{359-370}.
\newblock
\begin{APACrefDOI} \doi{10.1086/151968} \end{APACrefDOI}
\PrintBackRefs{\CurrentBib}

\bibitem [\protect \citeauthoryear {%
Bruzzone%
\ \protect \BOthers {.}}{%
Bruzzone%
\ \protect \BOthers {.}}{%
{\protect \APACyear {2015}}%
}]{%
JUICE-overview}
\APACinsertmetastar {%
JUICE-overview}%
\begin{APACrefauthors}%
Bruzzone, L.%
, Plaut, J\BPBI J.%
, Alberti, G.%
, Blankenship, D\BPBI D.%
, Bovolo, F.%
, Campbell, B\BPBI A.%
\BDBL {}others%
\end{APACrefauthors}%
\unskip\
\newblock
\APACrefYearMonthDay{2015}{}{}.
\newblock
{\BBOQ}\APACrefatitle {Jupiter ICY moon explorer (JUICE): Advances in the design of the radar for Icy Moons (RIME)} {Jupiter icy moon explorer (juice): Advances in the design of the radar for icy moons (rime)}.{\BBCQ}
\newblock
\BIn{} \APACrefbtitle {2015 IEEE international geoscience and remote sensing symposium (IGARSS)} {2015 ieee international geoscience and remote sensing symposium (igarss)}\ (\BPGS\ 1257--1260).
\PrintBackRefs{\CurrentBib}

\bibitem [\protect \citeauthoryear {%
Burke%
, Poggio%
, Logan%
\BCBL {}\ \BBA {} Rockway%
}{%
Burke%
\ \protect \BOthers {.}}{%
{\protect \APACyear {1979}}%
}]{%
NEC2}
\APACinsertmetastar {%
NEC2}%
\begin{APACrefauthors}%
Burke, G\BPBI J.%
, Poggio, A.%
, Logan, J.%
\BCBL {}\ \BBA {} Rockway, J.%
\end{APACrefauthors}%
\unskip\
\newblock
\APACrefYearMonthDay{1979}{}{}.
\newblock
{\BBOQ}\APACrefatitle {Numerical electromagnetic code (NEC)} {Numerical electromagnetic code (nec)}.{\BBCQ}
\newblock
\BIn{} \APACrefbtitle {1979 IEEE International Symposium on Electromagnetic Compatibility} {1979 ieee international symposium on electromagnetic compatibility}\ (\BPGS\ 1--3).
\PrintBackRefs{\CurrentBib}

\bibitem [\protect \citeauthoryear {%
{Cane}%
}{%
{Cane}%
}{%
{\protect \APACyear {1975}}%
}]{%
1975PASA....2..330C}
\APACinsertmetastar {%
1975PASA....2..330C}%
\begin{APACrefauthors}%
{Cane}, H\BPBI V.%
\end{APACrefauthors}%
\unskip\
\newblock
\APACrefYearMonthDay{1975}{{\APACmonth{10}}}{}.
\newblock
{\BBOQ}\APACrefatitle {{Low Frequency Maps of the Galaxy}} {{Low Frequency Maps of the Galaxy}}.{\BBCQ}
\newblock
\APACjournalVolNumPages{Publications of the Astronomical Society of Australia}{2}{6}{330-331}.
\newblock
\begin{APACrefDOI} \doi{10.1017/S1323358000014156} \end{APACrefDOI}
\PrintBackRefs{\CurrentBib}

\bibitem [\protect \citeauthoryear {%
{Cane}%
}{%
{Cane}%
}{%
{\protect \APACyear {1979}}%
}]{%
1979MNRAS.189..465C}
\APACinsertmetastar {%
1979MNRAS.189..465C}%
\begin{APACrefauthors}%
{Cane}, H\BPBI V.%
\end{APACrefauthors}%
\unskip\
\newblock
\APACrefYearMonthDay{1979}{{\APACmonth{11}}}{}.
\newblock
{\BBOQ}\APACrefatitle {{Spectra of the non-thermal radio radiation from the galactic polar regions.}} {{Spectra of the non-thermal radio radiation from the galactic polar regions.}}{\BBCQ}
\newblock
\APACjournalVolNumPages{Monthly Notices of the Royal Astronomical Society}{189}{}{465-478}.
\PrintBackRefs{\CurrentBib}

\bibitem [\protect \citeauthoryear {%
Cane%
}{%
Cane%
}{%
{\protect \APACyear {1979}}%
}]{%
cane-1979}
\APACinsertmetastar {%
cane-1979}%
\begin{APACrefauthors}%
Cane, H\BPBI V.%
\end{APACrefauthors}%
\unskip\
\newblock
\APACrefYearMonthDay{1979}{12}{}.
\newblock
{\BBOQ}\APACrefatitle {{Spectra of the non-thermal radio radiation from the galactic polar regions}} {{Spectra of the non-thermal radio radiation from the galactic polar regions}}.{\BBCQ}
\newblock
\APACjournalVolNumPages{Monthly Notices of the Royal Astronomical Society}{189}{3}{465-478}.
\newblock
\begin{APACrefURL} \url{https://doi.org/10.1093/mnras/189.3.465} \end{APACrefURL}
\newblock
\begin{APACrefDOI} \doi{10.1093/mnras/189.3.465} \end{APACrefDOI}
\PrintBackRefs{\CurrentBib}

\bibitem [\protect \citeauthoryear {%
{Caswell}%
}{%
{Caswell}%
}{%
{\protect \APACyear {1976}}%
}]{%
1976MNRAS.177..601C}
\APACinsertmetastar {%
1976MNRAS.177..601C}%
\begin{APACrefauthors}%
{Caswell}, J\BPBI L.%
\end{APACrefauthors}%
\unskip\
\newblock
\APACrefYearMonthDay{1976}{{\APACmonth{12}}}{}.
\newblock
{\BBOQ}\APACrefatitle {{A map of the northern sky at 10 MHz.}} {{A map of the northern sky at 10 MHz.}}{\BBCQ}
\newblock
\APACjournalVolNumPages{Monthly Notices of the Royal Astronomical Society}{177}{}{601-616}.
\newblock
\begin{APACrefDOI} \doi{10.1093/mnras/177.3.601} \end{APACrefDOI}
\PrintBackRefs{\CurrentBib}

\bibitem [\protect \citeauthoryear {%
Cong%
\ \protect \BOthers {.}}{%
Cong%
\ \protect \BOthers {.}}{%
{\protect \APACyear {2021}}%
}]{%
ULSA}
\APACinsertmetastar {%
ULSA}%
\begin{APACrefauthors}%
Cong, Y.%
, Yue, B.%
, Xu, Y.%
, Huang, Q.%
, Zuo, S.%
\BCBL {}\ \BBA {} Chen, X.%
\end{APACrefauthors}%
\unskip\
\newblock
\APACrefYearMonthDay{2021}{}{}.
\newblock
{\BBOQ}\APACrefatitle {An Ultralong-wavelength Sky Model with Absorption Effect} {An ultralong-wavelength sky model with absorption effect}.{\BBCQ}
\newblock
\APACjournalVolNumPages{The Astrophysical Journal}{914}{2}{128}.
\PrintBackRefs{\CurrentBib}

\bibitem [\protect \citeauthoryear {%
Cordes%
\ \BBA {} Lazio%
}{%
Cordes%
\ \BBA {} Lazio%
}{%
{\protect \APACyear {2002}}%
}]{%
NE2001}
\APACinsertmetastar {%
NE2001}%
\begin{APACrefauthors}%
Cordes, J\BPBI M.%
\BCBT {}\ \BBA {} Lazio, T\BPBI J\BPBI W.%
\end{APACrefauthors}%
\unskip\
\newblock
\APACrefYearMonthDay{2002}{}{}.
\newblock
{\BBOQ}\APACrefatitle {NE2001. I. A new model for the galactic distribution of free electrons and its fluctuations} {Ne2001. i. a new model for the galactic distribution of free electrons and its fluctuations}.{\BBCQ}
\newblock
\APACjournalVolNumPages{arXiv preprint astro-ph/0207156}{}{}{}.
\PrintBackRefs{\CurrentBib}

\bibitem [\protect \citeauthoryear {%
de Oliveira-Costa%
\ \protect \BOthers {.}}{%
de Oliveira-Costa%
\ \protect \BOthers {.}}{%
{\protect \APACyear {2008}}%
}]{%
GSM2008}
\APACinsertmetastar {%
GSM2008}%
\begin{APACrefauthors}%
de Oliveira-Costa, A.%
, Tegmark, M.%
, Gaensler, B.%
, Jonas, J.%
, Landecker, T.%
\BCBL {}\ \BBA {} Reich, P.%
\end{APACrefauthors}%
\unskip\
\newblock
\APACrefYearMonthDay{2008}{}{}.
\newblock
{\BBOQ}\APACrefatitle {A model of diffuse Galactic radio emission from 10 MHz to 100 GHz} {A model of diffuse galactic radio emission from 10 mhz to 100 ghz}.{\BBCQ}
\newblock
\APACjournalVolNumPages{Monthly Notices of the Royal Astronomical Society}{388}{1}{247--260}.
\PrintBackRefs{\CurrentBib}

\bibitem [\protect \citeauthoryear {%
Dowell%
\ \BBA {} Taylor%
}{%
Dowell%
\ \BBA {} Taylor%
}{%
{\protect \APACyear {2018}}%
}]{%
Dowell_2018}
\APACinsertmetastar {%
Dowell_2018}%
\begin{APACrefauthors}%
Dowell, J.%
\BCBT {}\ \BBA {} Taylor, G\BPBI B.%
\end{APACrefauthors}%
\unskip\
\newblock
\APACrefYearMonthDay{2018}{{\APACmonth{05}}}{}.
\newblock
{\BBOQ}\APACrefatitle {The Radio Background below 100 MHz} {The radio background below 100 mhz}.{\BBCQ}
\newblock
\APACjournalVolNumPages{The Astrophysical Journal Letters}{858}{1}{L9}.
\newblock
\begin{APACrefURL} \url{http://dx.doi.org/10.3847/2041-8213/aabf86} \end{APACrefURL}
\newblock
\begin{APACrefDOI} \doi{10.3847/2041-8213/aabf86} \end{APACrefDOI}
\PrintBackRefs{\CurrentBib}

\bibitem [\protect \citeauthoryear {%
Dowell%
, Taylor%
, Schinzel%
, Kassim%
\BCBL {}\ \BBA {} Stovall%
}{%
Dowell%
\ \protect \BOthers {.}}{%
{\protect \APACyear {2017}}%
}]{%
LFSM}
\APACinsertmetastar {%
LFSM}%
\begin{APACrefauthors}%
Dowell, J.%
, Taylor, G\BPBI B.%
, Schinzel, F\BPBI K.%
, Kassim, N\BPBI E.%
\BCBL {}\ \BBA {} Stovall, K.%
\end{APACrefauthors}%
\unskip\
\newblock
\APACrefYearMonthDay{2017}{}{}.
\newblock
{\BBOQ}\APACrefatitle {The LWA1 low frequency sky survey} {The lwa1 low frequency sky survey}.{\BBCQ}
\newblock
\APACjournalVolNumPages{Monthly Notices of the Royal Astronomical Society}{469}{4}{4537--4550}.
\PrintBackRefs{\CurrentBib}

\bibitem [\protect \citeauthoryear {%
{Dulk}%
, {Erickson}%
, {Manning}%
\BCBL {}\ \BBA {} {Bougeret}%
}{%
{Dulk}%
\ \protect \BOthers {.}}{%
{\protect \APACyear {2001}}%
}]{%
Dulk_2001}
\APACinsertmetastar {%
Dulk_2001}%
\begin{APACrefauthors}%
{Dulk}, G\BPBI A.%
, {Erickson}, W\BPBI C.%
, {Manning}, R.%
\BCBL {}\ \BBA {} {Bougeret}, J\BPBI L.%
\end{APACrefauthors}%
\unskip\
\newblock
\APACrefYearMonthDay{2001}{{\APACmonth{01}}}{}.
\newblock
{\BBOQ}\APACrefatitle {{Calibration of low-frequency radio telescopes using the galactic background radiation}} {{Calibration of low-frequency radio telescopes using the galactic background radiation}}.{\BBCQ}
\newblock
\APACjournalVolNumPages{Astronomy \& Astrophysics}{365}{}{294-300}.
\newblock
\begin{APACrefDOI} \doi{10.1051/0004-6361:20000006} \end{APACrefDOI}
\PrintBackRefs{\CurrentBib}

\bibitem [\protect \citeauthoryear {%
{Ellis}%
}{%
{Ellis}%
}{%
{\protect \APACyear {1964}}%
}]{%
1964Natur.204..171E}
\APACinsertmetastar {%
1964Natur.204..171E}%
\begin{APACrefauthors}%
{Ellis}, G\BPBI R\BPBI A.%
\end{APACrefauthors}%
\unskip\
\newblock
\APACrefYearMonthDay{1964}{{\APACmonth{10}}}{}.
\newblock
{\BBOQ}\APACrefatitle {{Spectra of the Galactic Radio Emissions}} {{Spectra of the Galactic Radio Emissions}}.{\BBCQ}
\newblock
\APACjournalVolNumPages{Nature}{204}{4954}{171-172}.
\newblock
\begin{APACrefDOI} \doi{10.1038/204171b0} \end{APACrefDOI}
\PrintBackRefs{\CurrentBib}

\bibitem [\protect \citeauthoryear {%
{Ellis}%
\ \BBA {} {Hamilton}%
}{%
{Ellis}%
\ \BBA {} {Hamilton}%
}{%
{\protect \APACyear {1966}}%
}]{%
1966ApJ...143..227E}
\APACinsertmetastar {%
1966ApJ...143..227E}%
\begin{APACrefauthors}%
{Ellis}, G\BPBI R\BPBI A.%
\BCBT {}\ \BBA {} {Hamilton}, P\BPBI A.%
\end{APACrefauthors}%
\unskip\
\newblock
\APACrefYearMonthDay{1966}{{\APACmonth{01}}}{}.
\newblock
{\BBOQ}\APACrefatitle {{Cosmic Radio Noise Survey at 4.7 Mc/s}} {{Cosmic Radio Noise Survey at 4.7 Mc/s}}.{\BBCQ}
\newblock
\APACjournalVolNumPages{The Astrophysical Journal}{143}{}{227}.
\newblock
\begin{APACrefDOI} \doi{10.1086/148493} \end{APACrefDOI}
\PrintBackRefs{\CurrentBib}

\bibitem [\protect \citeauthoryear {%
{Farrell}%
, {Desch}%
\BCBL {}\ \BBA {} {Zarka}%
}{%
{Farrell}%
\ \protect \BOthers {.}}{%
{\protect \APACyear {1999}}%
}]{%
1999JGR...10414025F}
\APACinsertmetastar {%
1999JGR...10414025F}%
\begin{APACrefauthors}%
{Farrell}, W\BPBI M.%
, {Desch}, M\BPBI D.%
\BCBL {}\ \BBA {} {Zarka}, P.%
\end{APACrefauthors}%
\unskip\
\newblock
\APACrefYearMonthDay{1999}{{\APACmonth{06}}}{}.
\newblock
{\BBOQ}\APACrefatitle {{On the possibility of coherent cyclotron emission from extrasolar planets}} {{On the possibility of coherent cyclotron emission from extrasolar planets}}.{\BBCQ}
\newblock
\APACjournalVolNumPages{jgr Planets}{104}{E6}{14025-14032}.
\newblock
\begin{APACrefDOI} \doi{10.1029/1998JE900050} \end{APACrefDOI}
\PrintBackRefs{\CurrentBib}

\bibitem [\protect \citeauthoryear {%
{Feenberg}%
\ \BBA {} {Primakoff}%
}{%
{Feenberg}%
\ \BBA {} {Primakoff}%
}{%
{\protect \APACyear {1948}}%
}]{%
1948PhRv...73..449F}
\APACinsertmetastar {%
1948PhRv...73..449F}%
\begin{APACrefauthors}%
{Feenberg}, E.%
\BCBT {}\ \BBA {} {Primakoff}, H.%
\end{APACrefauthors}%
\unskip\
\newblock
\APACrefYearMonthDay{1948}{{\APACmonth{03}}}{}.
\newblock
{\BBOQ}\APACrefatitle {{Interaction of Cosmic-Ray Primaries with Sunlight and Starlight}} {{Interaction of Cosmic-Ray Primaries with Sunlight and Starlight}}.{\BBCQ}
\newblock
\APACjournalVolNumPages{Physical Review}{73}{5}{449-469}.
\newblock
\begin{APACrefDOI} \doi{10.1103/PhysRev.73.449} \end{APACrefDOI}
\PrintBackRefs{\CurrentBib}

\bibitem [\protect \citeauthoryear {%
{Felten}%
\ \BBA {} {Morrison}%
}{%
{Felten}%
\ \BBA {} {Morrison}%
}{%
{\protect \APACyear {1966}}%
}]{%
1966ApJ...146..686F}
\APACinsertmetastar {%
1966ApJ...146..686F}%
\begin{APACrefauthors}%
{Felten}, J\BPBI E.%
\BCBT {}\ \BBA {} {Morrison}, P.%
\end{APACrefauthors}%
\unskip\
\newblock
\APACrefYearMonthDay{1966}{{\APACmonth{12}}}{}.
\newblock
{\BBOQ}\APACrefatitle {{Omnidirectional Inverse Compton and Synchrotron Radiation from Cosmic Distributions of Fast Electrons and Thermal Photons}} {{Omnidirectional Inverse Compton and Synchrotron Radiation from Cosmic Distributions of Fast Electrons and Thermal Photons}}.{\BBCQ}
\newblock
\APACjournalVolNumPages{The Astrophysical Journal}{146}{}{686}.
\newblock
\begin{APACrefDOI} \doi{10.1086/148946} \end{APACrefDOI}
\PrintBackRefs{\CurrentBib}

\bibitem [\protect \citeauthoryear {%
{Feretti}%
\ \BBA {} {Giovannini}%
}{%
{Feretti}%
\ \BBA {} {Giovannini}%
}{%
{\protect \APACyear {1996}}%
}]{%
1996IAUS..175..333F}
\APACinsertmetastar {%
1996IAUS..175..333F}%
\begin{APACrefauthors}%
{Feretti}, L.%
\BCBT {}\ \BBA {} {Giovannini}, G.%
\end{APACrefauthors}%
\unskip\
\newblock
\APACrefYearMonthDay{1996}{{\APACmonth{01}}}{}.
\newblock
{\BBOQ}\APACrefatitle {{Diffuse Cluster Radio Sources (Review)}} {{Diffuse Cluster Radio Sources (Review)}}.{\BBCQ}
\newblock
\BIn{} R\BPBI D.~{Ekers}, C.~{Fanti}\BCBL {}\ \BBA {} L.~{Padrielli}\ (\BEDS), \APACrefbtitle {Extragalactic Radio Sources} {Extragalactic radio sources}\ (\BVOL~175, \BPG~333).
\PrintBackRefs{\CurrentBib}

\bibitem [\protect \citeauthoryear {%
{Fixsen}%
\ \protect \BOthers {.}}{%
{Fixsen}%
\ \protect \BOthers {.}}{%
{\protect \APACyear {1994}}%
}]{%
FIRAS_Calibration}
\APACinsertmetastar {%
FIRAS_Calibration}%
\begin{APACrefauthors}%
{Fixsen}, D\BPBI J.%
, {Cheng}, E\BPBI S.%
, {Cottingham}, D\BPBI A.%
, {Eplee}, J., R.~E.%
, {Hewagama}, T.%
, {Isaacman}, R\BPBI B.%
\BDBL {}{Wright}, E\BPBI L.%
\end{APACrefauthors}%
\unskip\
\newblock
\APACrefYearMonthDay{1994}{{\APACmonth{01}}}{}.
\newblock
{\BBOQ}\APACrefatitle {{Calibration of the COBE FIRAS Instrument}} {{Calibration of the COBE FIRAS Instrument}}.{\BBCQ}
\newblock
\APACjournalVolNumPages{The Astophys. J.}{420}{}{457}.
\newblock
\begin{APACrefDOI} \doi{10.1086/173577} \end{APACrefDOI}
\PrintBackRefs{\CurrentBib}

\bibitem [\protect \citeauthoryear {%
Fixsen%
\ \protect \BOthers {.}}{%
Fixsen%
\ \protect \BOthers {.}}{%
{\protect \APACyear {1996}}%
}]{%
cobe-firas-main}
\APACinsertmetastar {%
cobe-firas-main}%
\begin{APACrefauthors}%
Fixsen, D\BPBI J.%
, Cheng, E\BPBI S.%
, Gales, J\BPBI M.%
, Mather, J\BPBI C.%
, Shafer, R\BPBI A.%
\BCBL {}\ \BBA {} Wright, E\BPBI L.%
\end{APACrefauthors}%
\unskip\
\newblock
\APACrefYearMonthDay{1996}{dec}{}.
\newblock
{\BBOQ}\APACrefatitle {The Cosmic Microwave Background Spectrum from the Full COBE* FIRAS Data Set} {The cosmic microwave background spectrum from the full cobe* firas data set}.{\BBCQ}
\newblock
\APACjournalVolNumPages{The Astophys. J.}{473}{2}{576}.
\newblock
\begin{APACrefURL} \url{https://dx.doi.org/10.1086/178173} \end{APACrefURL}
\newblock
\begin{APACrefDOI} \doi{10.1086/178173} \end{APACrefDOI}
\PrintBackRefs{\CurrentBib}

\bibitem [\protect \citeauthoryear {%
{Furlanetto}%
, {Oh}%
\BCBL {}\ \BBA {} {Briggs}%
}{%
{Furlanetto}%
\ \protect \BOthers {.}}{%
{\protect \APACyear {2006}}%
}]{%
2006PhR...433..181F}
\APACinsertmetastar {%
2006PhR...433..181F}%
\begin{APACrefauthors}%
{Furlanetto}, S\BPBI R.%
, {Oh}, S\BPBI P.%
\BCBL {}\ \BBA {} {Briggs}, F\BPBI H.%
\end{APACrefauthors}%
\unskip\
\newblock
\APACrefYearMonthDay{2006}{{\APACmonth{10}}}{}.
\newblock
{\BBOQ}\APACrefatitle {{Cosmology at low frequencies: The 21~cm transition and the high-redshift Universe}} {{Cosmology at low frequencies: The 21~cm transition and the high-redshift Universe}}.{\BBCQ}
\newblock
\APACjournalVolNumPages{physrep}{433}{4-6}{181-301}.
\newblock
\begin{APACrefDOI} \doi{10.1016/j.physrep.2006.08.002} \end{APACrefDOI}
\PrintBackRefs{\CurrentBib}

\bibitem [\protect \citeauthoryear {%
{Gurnett}%
\ \protect \BOthers {.}}{%
{Gurnett}%
\ \protect \BOthers {.}}{%
{\protect \APACyear {2004}}%
}]{%
2004SSRv..114..395G}
\APACinsertmetastar {%
2004SSRv..114..395G}%
\begin{APACrefauthors}%
{Gurnett}, D\BPBI A.%
, {Kurth}, W\BPBI S.%
, {Kirchner}, D\BPBI L.%
, {Hospodarsky}, G\BPBI B.%
, {Averkamp}, T\BPBI F.%
, {Zarka}, P.%
\BDBL {}{Pedersen}, A.%
\end{APACrefauthors}%
\unskip\
\newblock
\APACrefYearMonthDay{2004}{{\APACmonth{09}}}{}.
\newblock
{\BBOQ}\APACrefatitle {{The Cassini Radio and Plasma Wave Investigation}} {{The Cassini Radio and Plasma Wave Investigation}}.{\BBCQ}
\newblock
\APACjournalVolNumPages{ssr}{114}{1-4}{395-463}.
\newblock
\begin{APACrefDOI} \doi{10.1007/s11214-004-1434-0} \end{APACrefDOI}
\PrintBackRefs{\CurrentBib}

\bibitem [\protect \citeauthoryear {%
{Gurnett}%
\ \protect \BOthers {.}}{%
{Gurnett}%
\ \protect \BOthers {.}}{%
{\protect \APACyear {1992}}%
}]{%
1992SSRv...60..341G}
\APACinsertmetastar {%
1992SSRv...60..341G}%
\begin{APACrefauthors}%
{Gurnett}, D\BPBI A.%
, {Kurth}, W\BPBI S.%
, {Shaw}, R\BPBI R.%
, {Roux}, A.%
, {Gendrin}, R.%
, {Kennel}, C\BPBI F.%
\BDBL {}{Shawhan}, S\BPBI D.%
\end{APACrefauthors}%
\unskip\
\newblock
\APACrefYearMonthDay{1992}{{\APACmonth{05}}}{}.
\newblock
{\BBOQ}\APACrefatitle {{The Galileo Plasma wave investigation}} {{The Galileo Plasma wave investigation}}.{\BBCQ}
\newblock
\APACjournalVolNumPages{ssr}{60}{1-4}{341-355}.
\newblock
\begin{APACrefDOI} \doi{10.1007/BF00216861} \end{APACrefDOI}
\PrintBackRefs{\CurrentBib}

\bibitem [\protect \citeauthoryear {%
{Haddock}%
\ \BBA {} {Graedel}%
}{%
{Haddock}%
\ \BBA {} {Graedel}%
}{%
{\protect \APACyear {1968}}%
}]{%
1968AJS....73Q..62H}
\APACinsertmetastar {%
1968AJS....73Q..62H}%
\begin{APACrefauthors}%
{Haddock}, F\BPBI T.%
\BCBT {}\ \BBA {} {Graedel}, T\BPBI E.%
\end{APACrefauthors}%
\unskip\
\newblock
\APACrefYearMonthDay{1968}{{\APACmonth{01}}}{}.
\newblock
{\BBOQ}\APACrefatitle {{Low-Frequency Dynamic Spectra of Solar Bursts Observed from OGO-IlI.}} {{Low-Frequency Dynamic Spectra of Solar Bursts Observed from OGO-IlI.}}{\BBCQ}
\newblock
\APACjournalVolNumPages{The Astronomical Journal Supplement}{73}{}{62}.
\PrintBackRefs{\CurrentBib}

\bibitem [\protect \citeauthoryear {%
{Haddock}%
\ \BBA {} {Graedel}%
}{%
{Haddock}%
\ \BBA {} {Graedel}%
}{%
{\protect \APACyear {1970}}%
}]{%
1970ApJ...160..293H}
\APACinsertmetastar {%
1970ApJ...160..293H}%
\begin{APACrefauthors}%
{Haddock}, F\BPBI T.%
\BCBT {}\ \BBA {} {Graedel}, T\BPBI E.%
\end{APACrefauthors}%
\unskip\
\newblock
\APACrefYearMonthDay{1970}{{\APACmonth{04}}}{}.
\newblock
{\BBOQ}\APACrefatitle {{Dynamic Spectra of Type III Solar Bursts from 4 TO 2 MHz Observed by Ogo-Iii}} {{Dynamic Spectra of Type III Solar Bursts from 4 TO 2 MHz Observed by Ogo-Iii}}.{\BBCQ}
\newblock
\APACjournalVolNumPages{The Astrophysical Journal}{160}{}{293}.
\newblock
\begin{APACrefDOI} \doi{10.1086/150426} \end{APACrefDOI}
\PrintBackRefs{\CurrentBib}

\bibitem [\protect \citeauthoryear {%
{Hartz}%
}{%
{Hartz}%
}{%
{\protect \APACyear {1964}}%
{\protect \APACexlab {{\protect \BCnt {1}}}}}]{%
1964AnAp...27..823H}
\APACinsertmetastar {%
1964AnAp...27..823H}%
\begin{APACrefauthors}%
{Hartz}, T\BPBI R.%
\end{APACrefauthors}%
\unskip\
\newblock
\APACrefYearMonthDay{1964{\protect \BCnt {1}}}{{\APACmonth{02}}}{}.
\newblock
{\BBOQ}\APACrefatitle {{Observations of the galactic radio emission between 1,5 and 10 MHz from the Alouette satellite}} {{Observations of the galactic radio emission between 1,5 and 10 MHz from the Alouette satellite}}.{\BBCQ}
\newblock
\APACjournalVolNumPages{Annales d'Astrophysique}{27}{}{823}.
\PrintBackRefs{\CurrentBib}

\bibitem [\protect \citeauthoryear {%
{Hartz}%
}{%
{Hartz}%
}{%
{\protect \APACyear {1964}}%
{\protect \APACexlab {{\protect \BCnt {2}}}}}]{%
1964Natur.203..173H}
\APACinsertmetastar {%
1964Natur.203..173H}%
\begin{APACrefauthors}%
{Hartz}, T\BPBI R.%
\end{APACrefauthors}%
\unskip\
\newblock
\APACrefYearMonthDay{1964{\protect \BCnt {2}}}{{\APACmonth{07}}}{}.
\newblock
{\BBOQ}\APACrefatitle {{Spectrum of the Galactic Radio Emission between 1.5 and 10 Mc/s as observed from a Satellite}} {{Spectrum of the Galactic Radio Emission between 1.5 and 10 Mc/s as observed from a Satellite}}.{\BBCQ}
\newblock
\APACjournalVolNumPages{Nature}{203}{4941}{173-175}.
\newblock
\begin{APACrefDOI} \doi{10.1038/203173a0} \end{APACrefDOI}
\PrintBackRefs{\CurrentBib}

\bibitem [\protect \citeauthoryear {%
{Hartz}%
}{%
{Hartz}%
}{%
{\protect \APACyear {1969}}%
}]{%
1969JRASC..63R..94H}
\APACinsertmetastar {%
1969JRASC..63R..94H}%
\begin{APACrefauthors}%
{Hartz}, T\BPBI R.%
\end{APACrefauthors}%
\unskip\
\newblock
\APACrefYearMonthDay{1969}{{\APACmonth{01}}}{}.
\newblock
{\BBOQ}\APACrefatitle {{Radio astronomical observations from the Alouette II satellite.}} {{Radio astronomical observations from the Alouette II satellite.}}{\BBCQ}
\newblock
\APACjournalVolNumPages{Journal of the Royal Astronomical Society of Canada}{63}{}{94-95}.
\PrintBackRefs{\CurrentBib}

\bibitem [\protect \citeauthoryear {%
Haslam%
, Salter%
, Stoffel%
\BCBL {}\ \BBA {} Wilson%
}{%
Haslam%
\ \protect \BOthers {.}}{%
{\protect \APACyear {1982}}%
}]{%
haslam408}
\APACinsertmetastar {%
haslam408}%
\begin{APACrefauthors}%
Haslam, C.%
, Salter, C.%
, Stoffel, H.%
\BCBL {}\ \BBA {} Wilson, W.%
\end{APACrefauthors}%
\unskip\
\newblock
\APACrefYearMonthDay{1982}{}{}.
\newblock
{\BBOQ}\APACrefatitle {A 408 MHz all-sky continuum survey} {A 408 mhz all-sky continuum survey}.{\BBCQ}
\newblock
\APACjournalVolNumPages{Astronomy and Astrophysics Supplement Series}{47}{}{1}.
\PrintBackRefs{\CurrentBib}

\bibitem [\protect \citeauthoryear {%
Kasper%
, Lazio%
, Romero-Wolf%
, Lux%
\BCBL {}\ \BBA {} Neilsen%
}{%
Kasper%
\ \protect \BOthers {.}}{%
{\protect \APACyear {2022}}%
{\protect \APACexlab {{\protect \BCnt {1}}}}}]{%
Kasper_2022}
\APACinsertmetastar {%
Kasper_2022}%
\begin{APACrefauthors}%
Kasper, J.%
, Lazio, T\BPBI J\BPBI W.%
, Romero-Wolf, A.%
, Lux, J\BPBI P.%
\BCBL {}\ \BBA {} Neilsen, T.%
\end{APACrefauthors}%
\unskip\
\newblock
\APACrefYearMonthDay{2022{\protect \BCnt {1}}}{}{}.
\newblock
{\BBOQ}\APACrefatitle {The Sun Radio Interferometer Space Experiment (SunRISE) Mission} {The sun radio interferometer space experiment (sunrise) mission}.{\BBCQ}
\newblock
\BIn{} \APACrefbtitle {2022 IEEE Aerospace Conference (AERO)} {2022 ieee aerospace conference (aero)}\ (\BPG~1-8).
\newblock
\begin{APACrefDOI} \doi{10.1109/AERO53065.2022.9843607} \end{APACrefDOI}
\PrintBackRefs{\CurrentBib}

\bibitem [\protect \citeauthoryear {%
Kasper%
, Lazio%
, Romero-Wolf%
, Lux%
\BCBL {}\ \BBA {} Neilsen%
}{%
Kasper%
\ \protect \BOthers {.}}{%
{\protect \APACyear {2022}}%
{\protect \APACexlab {{\protect \BCnt {2}}}}}]{%
sunrise-overview-2022}
\APACinsertmetastar {%
sunrise-overview-2022}%
\begin{APACrefauthors}%
Kasper, J.%
, Lazio, T\BPBI J\BPBI W.%
, Romero-Wolf, A.%
, Lux, J\BPBI P.%
\BCBL {}\ \BBA {} Neilsen, T.%
\end{APACrefauthors}%
\unskip\
\newblock
\APACrefYearMonthDay{2022{\protect \BCnt {2}}}{}{}.
\newblock
{\BBOQ}\APACrefatitle {The Sun Radio Interferometer Space Experiment (SunRISE) Mission} {The sun radio interferometer space experiment (sunrise) mission}.{\BBCQ}
\newblock
\APACjournalVolNumPages{IEEE Aerospace Conference Proceedings}{2022-March}{}{}.
\newblock
\begin{APACrefDOI} \doi{10.1109/AERO53065.2022.9843607} \end{APACrefDOI}
\PrintBackRefs{\CurrentBib}

\bibitem [\protect \citeauthoryear {%
{Knoll}%
\ \protect \BOthers {.}}{%
{Knoll}%
\ \protect \BOthers {.}}{%
{\protect \APACyear {1978}}%
}]{%
1978ITGE...16..199K}
\APACinsertmetastar {%
1978ITGE...16..199K}%
\begin{APACrefauthors}%
{Knoll}, R.%
, {Epstein}, G.%
, {Hoang}, S.%
, {Huntzinger}, G.%
, {Steinberg}, J\BPBI L.%
, {Fainberg}, J.%
\BDBL {}{Stone}, R\BPBI G.%
\end{APACrefauthors}%
\unskip\
\newblock
\APACrefYearMonthDay{1978}{{\APACmonth{07}}}{}.
\newblock
{\BBOQ}\APACrefatitle {{The 3-dimensional radio mapping experiment (SBH) on ISEE-C.}} {{The 3-dimensional radio mapping experiment (SBH) on ISEE-C.}}{\BBCQ}
\newblock
\APACjournalVolNumPages{IEEE Transactions on Geoscience Electronics}{16}{}{199-204}.
\newblock
\begin{APACrefDOI} \doi{10.1109/TGE.1978.294546} \end{APACrefDOI}
\PrintBackRefs{\CurrentBib}

\bibitem [\protect \citeauthoryear {%
Kurth%
\ \protect \BOthers {.}}{%
Kurth%
\ \protect \BOthers {.}}{%
{\protect \APACyear {2017}}%
}]{%
juno-overview}
\APACinsertmetastar {%
juno-overview}%
\begin{APACrefauthors}%
Kurth, W.%
, Hospodarsky, G.%
, Kirchner, D.%
, Mokrzycki, B.%
, Averkamp, T.%
, Robison, W.%
\BDBL {}Zarka, P.%
\end{APACrefauthors}%
\unskip\
\newblock
\APACrefYearMonthDay{2017}{}{}.
\newblock
{\BBOQ}\APACrefatitle {The Juno waves investigation} {The juno waves investigation}.{\BBCQ}
\newblock
\APACjournalVolNumPages{Space Science Reviews}{213}{}{347--392}.
\PrintBackRefs{\CurrentBib}

\bibitem [\protect \citeauthoryear {%
{Kurth}%
\ \protect \BOthers {.}}{%
{Kurth}%
\ \protect \BOthers {.}}{%
{\protect \APACyear {2017}}%
}]{%
2017SSRv..213..347K}
\APACinsertmetastar {%
2017SSRv..213..347K}%
\begin{APACrefauthors}%
{Kurth}, W\BPBI S.%
, {Hospodarsky}, G\BPBI B.%
, {Kirchner}, D\BPBI L.%
, {Mokrzycki}, B\BPBI T.%
, {Averkamp}, T\BPBI F.%
, {Robison}, W\BPBI T.%
\BDBL {}{Zarka}, P.%
\end{APACrefauthors}%
\unskip\
\newblock
\APACrefYearMonthDay{2017}{{\APACmonth{11}}}{}.
\newblock
{\BBOQ}\APACrefatitle {{The Juno Waves Investigation}} {{The Juno Waves Investigation}}.{\BBCQ}
\newblock
\APACjournalVolNumPages{ssr}{213}{1-4}{347-392}.
\newblock
\begin{APACrefDOI} \doi{10.1007/s11214-017-0396-y} \end{APACrefDOI}
\PrintBackRefs{\CurrentBib}

\bibitem [\protect \citeauthoryear {%
{Lazio}%
\ \BBA {} {Farrell}%
}{%
{Lazio}%
\ \BBA {} {Farrell}%
}{%
{\protect \APACyear {2006}}%
}]{%
2006pre6.conf..603L}
\APACinsertmetastar {%
2006pre6.conf..603L}%
\begin{APACrefauthors}%
{Lazio}, T\BPBI J\BPBI W.%
\BCBT {}\ \BBA {} {Farrell}, W\BPBI M.%
\end{APACrefauthors}%
\unskip\
\newblock
\APACrefYearMonthDay{2006}{{\APACmonth{01}}}{}.
\newblock
{\BBOQ}\APACrefatitle {{Radio Detection of Extrasolar Planets: Present and Future Prospects}} {{Radio Detection of Extrasolar Planets: Present and Future Prospects}}.{\BBCQ}
\newblock
\BIn{} \APACrefbtitle {Planetary Radio Emissions VI} {Planetary radio emissions vi}\ (\BPG~603-610).
\PrintBackRefs{\CurrentBib}

\bibitem [\protect \citeauthoryear {%
{Liu}%
, {Pritchard}%
, {Tegmark}%
\BCBL {}\ \BBA {} {Loeb}%
}{%
{Liu}%
\ \protect \BOthers {.}}{%
{\protect \APACyear {2013}}%
}]{%
2013PhRvD..87d3002L}
\APACinsertmetastar {%
2013PhRvD..87d3002L}%
\begin{APACrefauthors}%
{Liu}, A.%
, {Pritchard}, J\BPBI R.%
, {Tegmark}, M.%
\BCBL {}\ \BBA {} {Loeb}, A.%
\end{APACrefauthors}%
\unskip\
\newblock
\APACrefYearMonthDay{2013}{{\APACmonth{02}}}{}.
\newblock
{\BBOQ}\APACrefatitle {{Global 21 cm signal experiments: A designer's guide}} {{Global 21 cm signal experiments: A designer's guide}}.{\BBCQ}
\newblock
\APACjournalVolNumPages{Phys.~Rev.~D}{87}{4}{043002}.
\newblock
\begin{APACrefDOI} \doi{10.1103/PhysRevD.87.043002} \end{APACrefDOI}
\PrintBackRefs{\CurrentBib}

\bibitem [\protect \citeauthoryear {%
{Ludwig}%
}{%
{Ludwig}%
}{%
{\protect \APACyear {1963}}%
}]{%
1963SSRv....2..175L}
\APACinsertmetastar {%
1963SSRv....2..175L}%
\begin{APACrefauthors}%
{Ludwig}, G\BPBI H.%
\end{APACrefauthors}%
\unskip\
\newblock
\APACrefYearMonthDay{1963}{{\APACmonth{08}}}{}.
\newblock
{\BBOQ}\APACrefatitle {{The Orbiting Geophysical Observatories}} {{The Orbiting Geophysical Observatories}}.{\BBCQ}
\newblock
\APACjournalVolNumPages{Scientific Studies of Reading}{2}{2}{175-218}.
\newblock
\begin{APACrefDOI} \doi{10.1007/BF00216779} \end{APACrefDOI}
\PrintBackRefs{\CurrentBib}

\bibitem [\protect \citeauthoryear {%
{Maksimovic}%
\ \protect \BOthers {.}}{%
{Maksimovic}%
\ \protect \BOthers {.}}{%
{\protect \APACyear {2020}}%
}]{%
2020A&A...642A..12M}
\APACinsertmetastar {%
2020A&A...642A..12M}%
\begin{APACrefauthors}%
{Maksimovic}, M.%
, {Bale}, S\BPBI D.%
, {Chust}, T.%
, {Khotyaintsev}, Y.%
, {Krasnoselskikh}, V.%
, {Kretzschmar}, M.%
\BDBL {}{Zouganelis}, I.%
\end{APACrefauthors}%
\unskip\
\newblock
\APACrefYearMonthDay{2020}{{\APACmonth{10}}}{}.
\newblock
{\BBOQ}\APACrefatitle {{The Solar Orbiter Radio and Plasma Waves (RPW) instrument}} {{The Solar Orbiter Radio and Plasma Waves (RPW) instrument}}.{\BBCQ}
\newblock
\APACjournalVolNumPages{aap}{642}{}{A12}.
\newblock
\begin{APACrefDOI} \doi{10.1051/0004-6361/201936214} \end{APACrefDOI}
\PrintBackRefs{\CurrentBib}

\bibitem [\protect \citeauthoryear {%
Malaspina%
\ \BBA {} Wilson~III%
}{%
Malaspina%
\ \BBA {} Wilson~III%
}{%
{\protect \APACyear {2016}}%
}]{%
dust_winds_2}
\APACinsertmetastar {%
dust_winds_2}%
\begin{APACrefauthors}%
Malaspina, D\BPBI M.%
\BCBT {}\ \BBA {} Wilson~III, L\BPBI B.%
\end{APACrefauthors}%
\unskip\
\newblock
\APACrefYearMonthDay{2016}{}{}.
\newblock
{\BBOQ}\APACrefatitle {A database of interplanetary and interstellar dust detected by the Wind spacecraft} {A database of interplanetary and interstellar dust detected by the wind spacecraft}.{\BBCQ}
\newblock
\APACjournalVolNumPages{Journal of Geophysical Research: Space Physics}{121}{10}{9369-9377}.
\newblock
\begin{APACrefURL} \url{https://agupubs.onlinelibrary.wiley.com/doi/abs/10.1002/2016JA023209} \end{APACrefURL}
\newblock
\begin{APACrefDOI} \doi{https://doi.org/10.1002/2016JA023209} \end{APACrefDOI}
\PrintBackRefs{\CurrentBib}

\bibitem [\protect \citeauthoryear {%
{Manning}%
\ \BBA {} {Dulk}%
}{%
{Manning}%
\ \BBA {} {Dulk}%
}{%
{\protect \APACyear {2001}}%
}]{%
2001A&A...372..663M}
\APACinsertmetastar {%
2001A&A...372..663M}%
\begin{APACrefauthors}%
{Manning}, R.%
\BCBT {}\ \BBA {} {Dulk}, G\BPBI A.%
\end{APACrefauthors}%
\unskip\
\newblock
\APACrefYearMonthDay{2001}{{\APACmonth{06}}}{}.
\newblock
{\BBOQ}\APACrefatitle {{The Galactic background radiation from 0.2 to 13.8 MHz}} {{The Galactic background radiation from 0.2 to 13.8 MHz}}.{\BBCQ}
\newblock
\APACjournalVolNumPages{aap}{372}{}{663-666}.
\newblock
\begin{APACrefDOI} \doi{10.1051/0004-6361:20010516} \end{APACrefDOI}
\PrintBackRefs{\CurrentBib}

\bibitem [\protect \citeauthoryear {%
Mather%
, Fixsen%
, Shafer%
, Mosier%
\BCBL {}\ \BBA {} Wilkinson%
}{%
Mather%
\ \protect \BOthers {.}}{%
{\protect \APACyear {1999}}%
}]{%
FIRAS_Calibration2}
\APACinsertmetastar {%
FIRAS_Calibration2}%
\begin{APACrefauthors}%
Mather, J\BPBI C.%
, Fixsen, D\BPBI J.%
, Shafer, R\BPBI A.%
, Mosier, C.%
\BCBL {}\ \BBA {} Wilkinson, D\BPBI T.%
\end{APACrefauthors}%
\unskip\
\newblock
\APACrefYearMonthDay{1999}{feb}{}.
\newblock
{\BBOQ}\APACrefatitle {Calibrator Design for the COBE* Far Infrared Absolute Spectrophotometer (FIRAS)} {Calibrator design for the cobe* far infrared absolute spectrophotometer (firas)}.{\BBCQ}
\newblock
\APACjournalVolNumPages{The Astrophysical Journal}{512}{2}{511}.
\newblock
\begin{APACrefURL} \url{https://dx.doi.org/10.1086/306805} \end{APACrefURL}
\newblock
\begin{APACrefDOI} \doi{10.1086/306805} \end{APACrefDOI}
\PrintBackRefs{\CurrentBib}

\bibitem [\protect \citeauthoryear {%
Meyer-Vernet%
\ \BBA {} Perche%
}{%
Meyer-Vernet%
\ \BBA {} Perche%
}{%
{\protect \APACyear {1989}}%
}]{%
Meyer-Vernet_1989}
\APACinsertmetastar {%
Meyer-Vernet_1989}%
\begin{APACrefauthors}%
Meyer-Vernet, N.%
\BCBT {}\ \BBA {} Perche, C.%
\end{APACrefauthors}%
\unskip\
\newblock
\APACrefYearMonthDay{1989}{}{}.
\newblock
{\BBOQ}\APACrefatitle {Tool kit for antennae and thermal noise near the plasma frequency} {Tool kit for antennae and thermal noise near the plasma frequency}.{\BBCQ}
\newblock
\APACjournalVolNumPages{Journal of Geophysical Research: Space Physics}{94}{A3}{2405--2415}.
\PrintBackRefs{\CurrentBib}

\bibitem [\protect \citeauthoryear {%
{Novaco}%
\ \BBA {} {Brown}%
}{%
{Novaco}%
\ \BBA {} {Brown}%
}{%
{\protect \APACyear {1978}}%
}]{%
Novaco_1978}
\APACinsertmetastar {%
Novaco_1978}%
\begin{APACrefauthors}%
{Novaco}, J\BPBI C.%
\BCBT {}\ \BBA {} {Brown}, L\BPBI W.%
\end{APACrefauthors}%
\unskip\
\newblock
\APACrefYearMonthDay{1978}{{\APACmonth{04}}}{}.
\newblock
{\BBOQ}\APACrefatitle {{Nonthermal galactic emission below 10 megahertz.}} {{Nonthermal galactic emission below 10 megahertz.}}{\BBCQ}
\newblock
\APACjournalVolNumPages{The Astrophysical Journal}{221}{}{114-123}.
\newblock
\begin{APACrefDOI} \doi{10.1086/156009} \end{APACrefDOI}
\PrintBackRefs{\CurrentBib}

\bibitem [\protect \citeauthoryear {%
Nyquist%
}{%
Nyquist%
}{%
{\protect \APACyear {1928}}%
}]{%
nyquist1928thermal}
\APACinsertmetastar {%
nyquist1928thermal}%
\begin{APACrefauthors}%
Nyquist, H.%
\end{APACrefauthors}%
\unskip\
\newblock
\APACrefYearMonthDay{1928}{}{}.
\newblock
{\BBOQ}\APACrefatitle {Thermal agitation of electric charge in conductors} {Thermal agitation of electric charge in conductors}.{\BBCQ}
\newblock
\APACjournalVolNumPages{Physical review}{32}{1}{110}.
\PrintBackRefs{\CurrentBib}

\bibitem [\protect \citeauthoryear {%
{Page}%
\ \protect \BOthers {.}}{%
{Page}%
\ \protect \BOthers {.}}{%
{\protect \APACyear {2022}}%
}]{%
Page_2022}
\APACinsertmetastar {%
Page_2022}%
\begin{APACrefauthors}%
{Page}, B.%
, {Bassett}, N.%
, {Lecacheux}, A.%
, {Pulupa}, M.%
, {Rapetti}, D.%
\BCBL {}\ \BBA {} {Bale}, S\BPBI D.%
\end{APACrefauthors}%
\unskip\
\newblock
\APACrefYearMonthDay{2022}{{\APACmonth{12}}}{}.
\newblock
{\BBOQ}\APACrefatitle {{The l = 2 spherical harmonic expansion coefficients of the sky brightness distribution between~0.5 and~7~MHz}} {{The l = 2 spherical harmonic expansion coefficients of the sky brightness distribution between~0.5 and~7~MHz}}.{\BBCQ}
\newblock
\APACjournalVolNumPages{aap}{668}{}{A127}.
\newblock
\begin{APACrefDOI} \doi{10.1051/0004-6361/202244621} \end{APACrefDOI}
\PrintBackRefs{\CurrentBib}

\bibitem [\protect \citeauthoryear {%
{Price}%
}{%
{Price}%
}{%
{\protect \APACyear {2016}}%
}]{%
pygdsm}
\APACinsertmetastar {%
pygdsm}%
\begin{APACrefauthors}%
{Price}, D\BPBI C.%
\end{APACrefauthors}%
\unskip\
\newblock
\APACrefYearMonthDay{2016}{{\APACmonth{03}}}{}.
\newblock
{\BBOQ}\APACrefatitle {{PyGDSM: Python interface to Global Diffuse Sky Models}} {{PyGDSM: Python interface to Global Diffuse Sky Models}}.{\BBCQ}
\newblock
\APAChowpublished {Astrophysics Source Code Library, record ascl:1603.013. [Software]}.
\PrintBackRefs{\CurrentBib}

\bibitem [\protect \citeauthoryear {%
{Pritchard}%
\ \BBA {} {Loeb}%
}{%
{Pritchard}%
\ \BBA {} {Loeb}%
}{%
{\protect \APACyear {2008}}%
}]{%
2008PhRvD..78j3511P}
\APACinsertmetastar {%
2008PhRvD..78j3511P}%
\begin{APACrefauthors}%
{Pritchard}, J\BPBI R.%
\BCBT {}\ \BBA {} {Loeb}, A.%
\end{APACrefauthors}%
\unskip\
\newblock
\APACrefYearMonthDay{2008}{{\APACmonth{11}}}{}.
\newblock
{\BBOQ}\APACrefatitle {{Evolution of the 21cm signal throughout cosmic history}} {{Evolution of the 21cm signal throughout cosmic history}}.{\BBCQ}
\newblock
\APACjournalVolNumPages{Phys.~Rev.~D}{78}{10}{103511}.
\newblock
\begin{APACrefDOI} \doi{10.1103/PhysRevD.78.103511} \end{APACrefDOI}
\PrintBackRefs{\CurrentBib}

\bibitem [\protect \citeauthoryear {%
{Pritchard}%
\ \BBA {} {Loeb}%
}{%
{Pritchard}%
\ \BBA {} {Loeb}%
}{%
{\protect \APACyear {2010}}%
}]{%
2010PhRvD..82b3006P}
\APACinsertmetastar {%
2010PhRvD..82b3006P}%
\begin{APACrefauthors}%
{Pritchard}, J\BPBI R.%
\BCBT {}\ \BBA {} {Loeb}, A.%
\end{APACrefauthors}%
\unskip\
\newblock
\APACrefYearMonthDay{2010}{{\APACmonth{07}}}{}.
\newblock
{\BBOQ}\APACrefatitle {{Constraining the unexplored period between the dark ages and reionization with observations of the global 21 cm signal}} {{Constraining the unexplored period between the dark ages and reionization with observations of the global 21 cm signal}}.{\BBCQ}
\newblock
\APACjournalVolNumPages{Phys.~Rev.~D}{82}{2}{023006}.
\newblock
\begin{APACrefDOI} \doi{10.1103/PhysRevD.82.023006} \end{APACrefDOI}
\PrintBackRefs{\CurrentBib}

\bibitem [\protect \citeauthoryear {%
{Rees}%
}{%
{Rees}%
}{%
{\protect \APACyear {1967}}%
}]{%
1967MNRAS.137..429R}
\APACinsertmetastar {%
1967MNRAS.137..429R}%
\begin{APACrefauthors}%
{Rees}, M\BPBI J.%
\end{APACrefauthors}%
\unskip\
\newblock
\APACrefYearMonthDay{1967}{{\APACmonth{01}}}{}.
\newblock
{\BBOQ}\APACrefatitle {{Studies in radio source structure-III. Inverse Compton radiation from radio sources}} {{Studies in radio source structure-III. Inverse Compton radiation from radio sources}}.{\BBCQ}
\newblock
\APACjournalVolNumPages{mnras}{137}{}{429}.
\newblock
\begin{APACrefDOI} \doi{10.1093/mnras/137.4.429} \end{APACrefDOI}
\PrintBackRefs{\CurrentBib}

\bibitem [\protect \citeauthoryear {%
{Reynolds}%
}{%
{Reynolds}%
}{%
{\protect \APACyear {1990}}%
}]{%
1990LNP...362..121R}
\APACinsertmetastar {%
1990LNP...362..121R}%
\begin{APACrefauthors}%
{Reynolds}, R\BPBI J.%
\end{APACrefauthors}%
\unskip\
\newblock
\APACrefYearMonthDay{1990}{}{}.
\newblock
{\BBOQ}\APACrefatitle {{The Low Density Ionized Component of the Interstellar Medium and Free-Free Absorption at High Galactic Latitudes}} {{The Low Density Ionized Component of the Interstellar Medium and Free-Free Absorption at High Galactic Latitudes}}.{\BBCQ}
\newblock
\BIn{} N\BPBI E.~{Kassim}\ \BBA {} K\BPBI W.~{Weiler}\ (\BEDS), \APACrefbtitle {Low Frequency Astrophysics from Space} {Low frequency astrophysics from space}\ (\BVOL~362, \BPG~121).
\newblock
\begin{APACrefDOI} \doi{10.1007/3-540-52891-1} \end{APACrefDOI}
\PrintBackRefs{\CurrentBib}

\bibitem [\protect \citeauthoryear {%
{Roger}%
, {Costain}%
, {Landecker}%
\BCBL {}\ \BBA {} {Swerdlyk}%
}{%
{Roger}%
\ \protect \BOthers {.}}{%
{\protect \APACyear {1999}}%
}]{%
Rogers_1999}
\APACinsertmetastar {%
Rogers_1999}%
\begin{APACrefauthors}%
{Roger}, R\BPBI S.%
, {Costain}, C\BPBI H.%
, {Landecker}, T\BPBI L.%
\BCBL {}\ \BBA {} {Swerdlyk}, C\BPBI M.%
\end{APACrefauthors}%
\unskip\
\newblock
\APACrefYearMonthDay{1999}{{\APACmonth{05}}}{}.
\newblock
{\BBOQ}\APACrefatitle {{The radio emission from the Galaxy at 22 MHz}} {{The radio emission from the Galaxy at 22 MHz}}.{\BBCQ}
\newblock
\APACjournalVolNumPages{Astronomy and Astrophysics Supplement}{137}{}{7-19}.
\newblock
\begin{APACrefDOI} \doi{10.1051/aas:1999239} \end{APACrefDOI}
\PrintBackRefs{\CurrentBib}

\bibitem [\protect \citeauthoryear {%
Rolla%
}{%
Rolla%
}{%
{\protect \APACyear {2024}}%
}]{%
zenodo-software}
\APACinsertmetastar {%
zenodo-software}%
\begin{APACrefauthors}%
Rolla, J.%
\end{APACrefauthors}%
\unskip\
\newblock
\APACrefYearMonthDay{2024}{}{}.
\newblock
{\BBOQ}\APACrefatitle {{An Instrument Error Budget for Space-Based Absolute Flux Measurements of the Sky Synchrotron Spectrum Below 20 MHz. Zenodo. [Software]}} {{An Instrument Error Budget for Space-Based Absolute Flux Measurements of the Sky Synchrotron Spectrum Below 20 MHz. Zenodo. [Software]}}.{\BBCQ}
\newblock

\newblock
\begin{APACrefDOI} \doi{10.5281/zenodo.11483379} \end{APACrefDOI}
\PrintBackRefs{\CurrentBib}

\bibitem [\protect \citeauthoryear {%
{Scarf}%
, {Fredricks}%
, {Gurnett}%
\BCBL {}\ \BBA {} {Smith}%
}{%
{Scarf}%
\ \protect \BOthers {.}}{%
{\protect \APACyear {1978}}%
}]{%
1978ITGE...16..191S}
\APACinsertmetastar {%
1978ITGE...16..191S}%
\begin{APACrefauthors}%
{Scarf}, F\BPBI L.%
, {Fredricks}, R\BPBI W.%
, {Gurnett}, D\BPBI A.%
\BCBL {}\ \BBA {} {Smith}, E\BPBI J.%
\end{APACrefauthors}%
\unskip\
\newblock
\APACrefYearMonthDay{1978}{{\APACmonth{07}}}{}.
\newblock
{\BBOQ}\APACrefatitle {{The ISEE-C plasma wave investigation.}} {{The ISEE-C plasma wave investigation.}}{\BBCQ}
\newblock
\APACjournalVolNumPages{IEEE Transactions on Geoscience Electronics}{16}{}{191-195}.
\newblock
\begin{APACrefDOI} \doi{10.1109/TGE.1978.294544} \end{APACrefDOI}
\PrintBackRefs{\CurrentBib}

\bibitem [\protect \citeauthoryear {%
Silk%
, Crawford%
, Elvis%
\BCBL {}\ \BBA {} Zarnecki%
}{%
Silk%
\ \protect \BOthers {.}}{%
{\protect \APACyear {2023}}%
}]{%
nature_overview}
\APACinsertmetastar {%
nature_overview}%
\begin{APACrefauthors}%
Silk, J.%
, Crawford, I.%
, Elvis, M.%
\BCBL {}\ \BBA {} Zarnecki, J.%
\end{APACrefauthors}%
\unskip\
\newblock
\APACrefYearMonthDay{2023}{}{}.
\newblock
{\BBOQ}\APACrefatitle {The next decades for astronomy from the Moon} {The next decades for astronomy from the moon}.{\BBCQ}
\newblock
\APACjournalVolNumPages{Nature Astronomy}{7}{6}{648--650}.
\PrintBackRefs{\CurrentBib}

\bibitem [\protect \citeauthoryear {%
{Singal}%
\ \protect \BOthers {.}}{%
{Singal}%
\ \protect \BOthers {.}}{%
{\protect \APACyear {2023}}%
}]{%
Singal_2023}
\APACinsertmetastar {%
Singal_2023}%
\begin{APACrefauthors}%
{Singal}, J.%
, {Fornengo}, N.%
, {Regis}, M.%
, {Bernardi}, G.%
, {Bordenave}, D.%
, {Branchini}, E.%
\BDBL {}{Todarello}, E.%
\end{APACrefauthors}%
\unskip\
\newblock
\APACrefYearMonthDay{2023}{}{}.
\newblock
{\BBOQ}\APACrefatitle {{The Second Radio Synchrotron Background Workshop: Conference Summary and Report}} {{The Second Radio Synchrotron Background Workshop: Conference Summary and Report}}.{\BBCQ}
\newblock
\APACjournalVolNumPages{Publ.\ Astron.\ Pacific}{135}{1045}{}.
\newblock
\begin{APACrefDOI} \doi{10.1088/1538-3873/acbdbf} \end{APACrefDOI}
\PrintBackRefs{\CurrentBib}

\bibitem [\protect \citeauthoryear {%
Spinelli%
\ \protect \BOthers {.}}{%
Spinelli%
\ \protect \BOthers {.}}{%
{\protect \APACyear {2021}}%
}]{%
Spinelli_2021}
\APACinsertmetastar {%
Spinelli_2021}%
\begin{APACrefauthors}%
Spinelli, M.%
, Bernardi, G.%
, Garsden, H.%
, Greenhill, L\BPBI J.%
, Fialkov, A.%
, Dowell, J.%
\BCBL {}\ \BBA {} Price, D\BPBI C.%
\end{APACrefauthors}%
\unskip\
\newblock
\APACrefYearMonthDay{2021}{05}{}.
\newblock
{\BBOQ}\APACrefatitle {{Spectral index of the Galactic foreground emission in the 50–87 MHz range}} {{Spectral index of the Galactic foreground emission in the 50–87 MHz range}}.{\BBCQ}
\newblock
\APACjournalVolNumPages{Monthly Notices of the Royal Astronomical Society}{505}{2}{1575-1588}.
\newblock
\begin{APACrefURL} \url{https://doi.org/10.1093/mnras/stab1363} \end{APACrefURL}
\newblock
\begin{APACrefDOI} \doi{10.1093/mnras/stab1363} \end{APACrefDOI}
\PrintBackRefs{\CurrentBib}

\bibitem [\protect \citeauthoryear {%
{Warwick}%
, {Pearce}%
, {Peltzer}%
\BCBL {}\ \BBA {} {Riddle}%
}{%
{Warwick}%
\ \protect \BOthers {.}}{%
{\protect \APACyear {1977}}%
}]{%
1977SSRv...21..309W}
\APACinsertmetastar {%
1977SSRv...21..309W}%
\begin{APACrefauthors}%
{Warwick}, J\BPBI W.%
, {Pearce}, J\BPBI B.%
, {Peltzer}, R\BPBI G.%
\BCBL {}\ \BBA {} {Riddle}, A\BPBI C.%
\end{APACrefauthors}%
\unskip\
\newblock
\APACrefYearMonthDay{1977}{{\APACmonth{12}}}{}.
\newblock
{\BBOQ}\APACrefatitle {{Planetary Radio Astronomy Experiment for Voyager Missions}} {{Planetary Radio Astronomy Experiment for Voyager Missions}}.{\BBCQ}
\newblock
\APACjournalVolNumPages{ssr}{21}{3}{309-327}.
\newblock
\begin{APACrefDOI} \doi{10.1007/BF00211544} \end{APACrefDOI}
\PrintBackRefs{\CurrentBib}

\bibitem [\protect \citeauthoryear {%
{Weber}%
}{%
{Weber}%
}{%
{\protect \APACyear {1975}}%
}]{%
1975RF.....19..250W}
\APACinsertmetastar {%
1975RF.....19..250W}%
\begin{APACrefauthors}%
{Weber}, R\BPBI R.%
\end{APACrefauthors}%
\unskip\
\newblock
\APACrefYearMonthDay{1975}{{\APACmonth{10}}}{}.
\newblock
{\BBOQ}\APACrefatitle {{The radio astronomy experiment on Helios A and B /E 5c/}} {{The radio astronomy experiment on Helios A and B /E 5c/}}.{\BBCQ}
\newblock
\APACjournalVolNumPages{Raumfahrtforschung}{19}{}{250-252}.
\PrintBackRefs{\CurrentBib}

\bibitem [\protect \citeauthoryear {%
{Weber}%
, {Alexander}%
\BCBL {}\ \BBA {} {Stone}%
}{%
{Weber}%
\ \protect \BOthers {.}}{%
{\protect \APACyear {1971}}%
{\protect \APACexlab {{\protect \BCnt {1}}}}}]{%
1971RaSc....6.1085W}
\APACinsertmetastar {%
1971RaSc....6.1085W}%
\begin{APACrefauthors}%
{Weber}, R\BPBI R.%
, {Alexander}, J\BPBI K.%
\BCBL {}\ \BBA {} {Stone}, R\BPBI G.%
\end{APACrefauthors}%
\unskip\
\newblock
\APACrefYearMonthDay{1971{\protect \BCnt {1}}}{{\APACmonth{12}}}{}.
\newblock
{\BBOQ}\APACrefatitle {{The Radio Astronomy Explorer Satellite, a low-frequency observatory}} {{The Radio Astronomy Explorer Satellite, a low-frequency observatory}}.{\BBCQ}
\newblock
\APACjournalVolNumPages{Radio Science}{6}{}{1085}.
\newblock
\begin{APACrefDOI} \doi{10.1029/RS006i012p01085} \end{APACrefDOI}
\PrintBackRefs{\CurrentBib}

\bibitem [\protect \citeauthoryear {%
{Weber}%
, {Alexander}%
\BCBL {}\ \BBA {} {Stone}%
}{%
{Weber}%
\ \protect \BOthers {.}}{%
{\protect \APACyear {1971}}%
{\protect \APACexlab {{\protect \BCnt {2}}}}}]{%
Weber_1971}
\APACinsertmetastar {%
Weber_1971}%
\begin{APACrefauthors}%
{Weber}, R\BPBI R.%
, {Alexander}, J\BPBI K.%
\BCBL {}\ \BBA {} {Stone}, R\BPBI G.%
\end{APACrefauthors}%
\unskip\
\newblock
\APACrefYearMonthDay{1971{\protect \BCnt {2}}}{{\APACmonth{12}}}{}.
\newblock
{\BBOQ}\APACrefatitle {{The Radio Astronomy Explorer Satellite, a low-frequency observatory}} {{The Radio Astronomy Explorer Satellite, a low-frequency observatory}}.{\BBCQ}
\newblock
\APACjournalVolNumPages{Radio Science}{6}{}{1085}.
\newblock
\begin{APACrefDOI} \doi{10.1029/RS006i012p01085} \end{APACrefDOI}
\PrintBackRefs{\CurrentBib}

\bibitem [\protect \citeauthoryear {%
Wilson~III%
\ \protect \BOthers {.}}{%
Wilson~III%
\ \protect \BOthers {.}}{%
{\protect \APACyear {2021}}%
}]{%
dust_impacts_winds}
\APACinsertmetastar {%
dust_impacts_winds}%
\begin{APACrefauthors}%
Wilson~III, L\BPBI B.%
, Brosius, A\BPBI L.%
, Gopalswamy, N.%
, Nieves-Chinchilla, T.%
, Szabo, A.%
, Hurley, K.%
\BDBL {}TenBarge, J\BPBI M.%
\end{APACrefauthors}%
\unskip\
\newblock
\APACrefYearMonthDay{2021}{}{}.
\newblock
{\BBOQ}\APACrefatitle {A Quarter Century of Wind Spacecraft Discoveries} {A quarter century of wind spacecraft discoveries}.{\BBCQ}
\newblock
\APACjournalVolNumPages{Reviews of Geophysics}{59}{2}{e2020RG000714}.
\newblock
\begin{APACrefURL} \url{https://agupubs.onlinelibrary.wiley.com/doi/abs/10.1029/2020RG000714} \end{APACrefURL}
\newblock
\APACrefnote{e2020RG000714 2020RG000714}
\newblock
\begin{APACrefDOI} \doi{https://doi.org/10.1029/2020RG000714} \end{APACrefDOI}
\PrintBackRefs{\CurrentBib}

\bibitem [\protect \citeauthoryear {%
Yanping-Cong%
}{%
Yanping-Cong%
}{%
{\protect \APACyear {2021}}%
}]{%
ulsa_code}
\APACinsertmetastar {%
ulsa_code}%
\begin{APACrefauthors}%
Yanping-Cong.%
\end{APACrefauthors}%
\unskip\
\newblock
\APACrefYearMonthDay{2021}{}{}.
\newblock
{\BBOQ}\APACrefatitle {{Yanping-Cong/ULSA. Zenodo. [Software]}} {{Yanping-Cong/ULSA. Zenodo. [Software]}}.{\BBCQ}
\newblock

\newblock
\begin{APACrefDOI} \doi{10.5281/zenodo.4663463} \end{APACrefDOI}
\PrintBackRefs{\CurrentBib}

\bibitem [\protect \citeauthoryear {%
{Yorks}%
}{%
{Yorks}%
}{%
{\protect \APACyear {1971}}%
}]{%
1971ITIM...20...86Y}
\APACinsertmetastar {%
1971ITIM...20...86Y}%
\begin{APACrefauthors}%
{Yorks}, R\BPBI G.%
\end{APACrefauthors}%
\unskip\
\newblock
\APACrefYearMonthDay{1971}{{\APACmonth{01}}}{}.
\newblock
{\BBOQ}\APACrefatitle {{Instrumentation for measurement of cosmic noise at 2.0 and 2.5 MHz from a polar orbiting geophysical observatory.}} {{Instrumentation for measurement of cosmic noise at 2.0 and 2.5 MHz from a polar orbiting geophysical observatory.}}{\BBCQ}
\newblock
\APACjournalVolNumPages{IEEE Transactions on Instrumentation Measurement}{20}{}{86-94}.
\newblock
\begin{APACrefDOI} \doi{10.1109/TIM.1971.5570699} \end{APACrefDOI}
\PrintBackRefs{\CurrentBib}

\bibitem [\protect \citeauthoryear {%
{Zarka}%
, {Cecconi}%
\BCBL {}\ \BBA {} {Kurth}%
}{%
{Zarka}%
\ \protect \BOthers {.}}{%
{\protect \APACyear {2004}}%
}]{%
Zarka_2004}
\APACinsertmetastar {%
Zarka_2004}%
\begin{APACrefauthors}%
{Zarka}, P.%
, {Cecconi}, B.%
\BCBL {}\ \BBA {} {Kurth}, W\BPBI S.%
\end{APACrefauthors}%
\unskip\
\newblock
\APACrefYearMonthDay{2004}{{\APACmonth{09}}}{}.
\newblock
{\BBOQ}\APACrefatitle {{Jupiter's low-frequency radio spectrum from Cassini/Radio and Plasma Wave Science (RPWS) absolute flux density measurements}} {{Jupiter's low-frequency radio spectrum from Cassini/Radio and Plasma Wave Science (RPWS) absolute flux density measurements}}.{\BBCQ}
\newblock
\APACjournalVolNumPages{Journal of Geophysical Research (Space Physics)}{109}{A9}{A09S15}.
\newblock
\begin{APACrefDOI} \doi{10.1029/2003JA010260} \end{APACrefDOI}
\PrintBackRefs{\CurrentBib}

\bibitem [\protect \citeauthoryear {%
{Zarka}%
, {Treumann}%
, {Ryabov}%
\BCBL {}\ \BBA {} {Ryabov}%
}{%
{Zarka}%
\ \protect \BOthers {.}}{%
{\protect \APACyear {2001}}%
}]{%
2001Ap&SS.277..293Z}
\APACinsertmetastar {%
2001Ap&SS.277..293Z}%
\begin{APACrefauthors}%
{Zarka}, P.%
, {Treumann}, R\BPBI A.%
, {Ryabov}, B\BPBI P.%
\BCBL {}\ \BBA {} {Ryabov}, V\BPBI B.%
\end{APACrefauthors}%
\unskip\
\newblock
\APACrefYearMonthDay{2001}{{\APACmonth{06}}}{}.
\newblock
{\BBOQ}\APACrefatitle {{Magnetically-Driven Planetary Radio Emissions and Application to Extrasolar Planets}} {{Magnetically-Driven Planetary Radio Emissions and Application to Extrasolar Planets}}.{\BBCQ}
\newblock
\APACjournalVolNumPages{apss}{277}{}{293-300}.
\newblock
\begin{APACrefDOI} \doi{10.1023/A:1012221527425} \end{APACrefDOI}
\PrintBackRefs{\CurrentBib}

\bibitem [\protect \citeauthoryear {%
{Zaslavsky}%
, {Meyer-Vernet}%
, {Hoang}%
, {Maksimovic}%
\BCBL {}\ \BBA {} {Bale}%
}{%
{Zaslavsky}%
\ \protect \BOthers {.}}{%
{\protect \APACyear {2011}}%
}]{%
Zaslavsky_2011RaSc}
\APACinsertmetastar {%
Zaslavsky_2011RaSc}%
\begin{APACrefauthors}%
{Zaslavsky}, A.%
, {Meyer-Vernet}, N.%
, {Hoang}, S.%
, {Maksimovic}, M.%
\BCBL {}\ \BBA {} {Bale}, S\BPBI D.%
\end{APACrefauthors}%
\unskip\
\newblock
\APACrefYearMonthDay{2011}{{\APACmonth{03}}}{}.
\newblock
{\BBOQ}\APACrefatitle {{On the antenna calibration of space radio instruments using the galactic background: General formulas and application to STEREO/WAVES}} {{On the antenna calibration of space radio instruments using the galactic background: General formulas and application to STEREO/WAVES}}.{\BBCQ}
\newblock
\APACjournalVolNumPages{Radio Science}{46}{2}{RS2008}.
\newblock
\begin{APACrefDOI} \doi{10.1029/2010RS004464} \end{APACrefDOI}
\PrintBackRefs{\CurrentBib}

\bibitem [\protect \citeauthoryear {%
Zheng%
\ \protect \BOthers {.}}{%
Zheng%
\ \protect \BOthers {.}}{%
{\protect \APACyear {2017}}%
}]{%
GSM2016}
\APACinsertmetastar {%
GSM2016}%
\begin{APACrefauthors}%
Zheng, H.%
, Tegmark, M.%
, Dillon, J\BPBI S.%
, Kim, D\BPBI A.%
, Liu, A.%
, Neben, A\BPBI R.%
\BDBL {}Reich, W.%
\end{APACrefauthors}%
\unskip\
\newblock
\APACrefYearMonthDay{2017}{}{}.
\newblock
{\BBOQ}\APACrefatitle {An improved model of diffuse galactic radio emission from 10 MHz to 5 THz} {An improved model of diffuse galactic radio emission from 10 mhz to 5 thz}.{\BBCQ}
\newblock
\APACjournalVolNumPages{Monthly Notices of the Royal Astronomical Society}{464}{3}{3486--3497}.
\PrintBackRefs{\CurrentBib}

\end{thebibliography}

%
%
%
%
%

\end{document}